\documentclass{aastex}

\usepackage{amssymb,amsmath,natbib,gensymb}
\usepackage{graphicx,graphics}
\usepackage{multirow}
\usepackage{subfigure}
\usepackage{color}
\usepackage[colorlinks,linkcolor=blue, anchorcolor=blue, citecolor=blue,filecolor=blue,menucolor=blue,runcolor=blue,urlcolor=blue]{hyperref}

\shorttitle{Gas Dynamics of a Luminous $z=6.13$ Quasar}
\shortauthors{Yali Shao, Ran Wang et al. 2017}

\newcommand{\cii}{[C\,{\sc ii}]}
\newcommand{\mgii}{Mg\,{\sc ii}}

\begin{document}

\title{Gas dynamics of a luminous $z$ = 6.13 quasar ULAS J1319$+$0950 revealed by ALMA high resolution observations}

\author{Yali Shao\altaffilmark{1,2,5}$^{,\dagger}$, Ran Wang\altaffilmark{2}$^{,\bigstar}$, Gareth C. Jones\altaffilmark{15,5}, Chris L. Carilli\altaffilmark{3,5}, Fabian Walter\altaffilmark{5,6,16}, Xiaohui Fan\altaffilmark{7}, Dominik A. Riechers\altaffilmark{8}, Frank Bertoldi\altaffilmark{9}, Jeff Wagg\altaffilmark{4}, Michael A. Strauss\altaffilmark{10}, Alain Omont\altaffilmark{12}, Pierre Cox\altaffilmark{11}, Linhua Jiang\altaffilmark{1,2}, Desika Narayanan\altaffilmark{14}, Karl M. Menten\altaffilmark{13}}

\altaffiltext{1}{Department of Astronomy, School of Physics, Peking University, Beijing 100871, China}
\altaffiltext{2}{Kavli Institute for Astronomy and Astrophysics, Peking University, Beijing 100871, China}
\altaffiltext{3}{Cavendish Laboratory, 19 J. J. Thomson Avenue, Cambridge CB3 0HE, UK}
\altaffiltext{4}{SKA Organization, Lower Withington Macclesfield, Cheshire SK11 9DL, UK}
\altaffiltext{5}{National Radio Astronomy Observatory, Socorro, NM 87801-0387, USA}
\altaffiltext{6}{Max-Planck-Institut for Astronomie, K\"{o}nigstuhl 17, D-69117 Heidelberg, Germany}
\altaffiltext{7}{Steward Observatory, University of Arizona, 933 North Cherry Avenue, Tucson, AZ 85721, USA}
\altaffiltext{8}{Department of Astronomy, Cornell University, 220 Space Sciences Building, Ithaca, NY 14853, USA}
\altaffiltext{9}{Argelander-Institut f\"{u}r Astronomie, University at Bonn, Auf dem H\"{u}gel 71, D-53121 Bonn, Germany}
\altaffiltext{10}{Department of Astrophysical Sciences, Princeton University, Princeton, NJ 08544, USA}
\altaffiltext{11}{Institute de Radioastronomie Millimetrique, St. Martin d'Heres, F-38406, France}
\altaffiltext{12}{Institut d'Astrophysique de Paris, CNRS and Universite Pierre et Marie Curie, Paris, France}
\altaffiltext{13}{Max-Planck-Institut fur Radioastronomie, Auf dem H\"{u}gel 69, 53121 Bonn, Germany}
\altaffiltext{14}{Haverford College, 370 W Lancaster Ave, Haverford, PA 19041, USA}
\altaffiltext{15}{Physics Department, New Mexico Institute of Mining and Technology, Socorro, NM 87801, USA}
\altaffiltext{16}{Astronomy Department, California Institute of Technology, MC105-24, Pasadena, CA 91125, USA}
\altaffiltext{$\dagger$}{E-mail: \href{mailto:yshao@nrao.edu}{yshao@nrao.edu}}
\altaffiltext{$\bigstar$}{E-mail: \href{mailto:rwangkiaa@pku.edu.cn}{rwangkiaa@pku.edu.cn}}

\begin{abstract}
We present new Atacama Large Millimeter/submillimeter Array (ALMA) observations of the dust continuum and [\ion{C}{2}] 158 $\mu$m fine structure line emission towards a far-infrared-luminous quasar, ULAS J131911.29$+$095051.4 at $z=6.13$, and combine the new Cycle 1 data with ALMA Cycle 0 data. The combined data have an angular resolution $\sim$ $0\farcs3$, and resolve both the dust continuum and the [\ion{C}{2}]  line emission on few kpc scales.  The [\ion{C}{2}] line emission is more irregular than the dust continuum emission which suggests different distributions between the dust and [\ion{C}{2}]-emitting gas. The combined data confirm the [\ion{C}{2}] velocity gradient that we previously detected in lower resolution ALMA image from Cycle 0 data alone. We apply a tilted ring model to the [\ion{C}{2}] velocity map to obtain a rotation curve, and constrain the circular velocity to be 427 $\pm$ 55 km s$^{-1}$ at a radius of 3.2 kpc with an inclination angle of 34 $\degr$. We measure the dynamical mass within the 3.2 kpc region to be 13.4$_{-5.3}^{+7.8}$ $\times 10^{10}\,M_{\odot}$. This yields a black hole and host galaxy mass ratio of 0.020$_{-0.007}^{+0.013}$, which is about 4$_{-2}^{+3}$ times higher than the present-day $M_{\rm BH}$/$M_{\rm bulge}$ ratio. This suggests that the supermassive black hole grows the bulk of its mass before the formation of the most of stellar mass in this quasar host galaxy in the early universe.
\end{abstract}

\keywords{galaxies: evolution --- galaxies: active --- galaxies: high-redshift --- submillimeter: galaxies --- quasars: general --- radio lines: galaxies }

\section{Introduction}
In recent years, more than 200 quasars at 5.7 $<$ $z$ $<$ 7.1 have been discovered in large optical and near-infrared surveys (e.g.,  \citealt{Fan2006lala}; \citealt{Jiang2015, Jiang2016}; \citealt{Venemans2007, Venemans2015b}; \citealt{Mortlock2009, Mortlock2011}; \citealt{Banados2016}; \citealt{Reed2017}; \citealt{Matsuoka2016}). Millimeter observations of the dust continuum and molecular CO indicate active star formation at rates of a few hundred to thousand $M_{\odot}$ yr$^{-1}$ in the host galaxies of about 30$\%$ of optically luminous quasars at $z \sim$ 6 (e.g., \citealt{Petric2003}; \citealt{Priddey2003}; \citealt{Bertoldi2003a, Bertoldi2003b}; \citealt{Wang2008thermal, Wang2011fir}). These quasar-starburst systems provide unique laboratories to study the formation of the first supermassive black holes (SMBHs) and their host galaxies close to the end of cosmic reionization. 

Bright [\ion{C}{2}]  158 $\mu$m fine structure line emission has been widely detected in high redshift quasar-starburst systems (\citealt{Maiolino2012}; \citealt{Wang2013}; \citealt{Willott2013,Willott2015}; \citealt{Venemans2016}). The [\ion{C}{2}]  line is one of the primary coolants of the star-forming interstellar medium (ISM). Thus, it directly traces the distribution of star formation activity and kinematic properties of the atomic/ionized gas in quasar host galaxies (\citealt{Kimball2015}; \citealt{DiazSantos2016}; \citealt{Venemans2017}). Sixteen quasars at $5.7 < z < 7.1$ are detected in [\ion{C}{2}]  line emission, with modern submm/mm interferometer arrays such as the NOrthern Extended Millimeter Array (NOEMA) and ALMA (\citealt{Walter2009}; \citealt{Wang2013, WangRan2016}; \citealt{Willott2013, Willott2015}; \citealt{Venemans2012, Venemans2016, Venemans2017}; \citealt{Banados2015}). These objects have [\ion{C}{2}] to far-infrared (FIR) luminosity ratios over a wide range of (0.19$-$4.8) $\times$ 10$^{-3}$ (\citealt{Walter2009}; \citealt{Willott2015}), indicating that the ISM is in a complex physical state powered by both AGN and star formation activity. The [\ion{C}{2}]  line emission in fourteen of these quasars have been observed with sub-arcsec resolution, and the inferred source sizes are 1.5$-$3.3 kpc (\citealt{Wang2013}; \citealt{Venemans2016, Venemans2017}; \citealt{Walter2009}; \citealt{Willott2013, Willott2015}). Six of them show clear velocity gradients (\citealt{Willott2013}; \citealt{Wang2013}), providing constraints on the dynamical mass. In these objects, the black hole to bulge mass ratio appears to be above the correlation defined by local objects \citep{Wang2013}. However, these studies were limited by the moderate angular resolution of the early ALMA observations (typically $0\farcs7$), resulting in a strong degeneracy between inclination angle and intrinsic rotational velocity.

In this paper, we report on ALMA Cycle 1 observations of a FIR-luminous quasar ULAS J131911.29$+$095051.4 (hereafter J1319$+$0950) at $z$ = 6.13, and combine it with ALMA Cycle 0 data to study gas dynamics. \citet{Mortlock2009} discovered this optically bright quasar from UKIRT Infrared Deep Sky Survey (UKIDSS) with $m_{1450\AA}$ = 19.65. \citet{Wang2011fir} observed this quasar by PdBI and measured the 250 GHz dust continuum emission using MAMBO, which suggests that it is a very FIR-luminous quasar. They also detected the 1.4 GHz radio continuum and the CO (6$-$5) line emission. The redshift measured from the CO (6$-$5) line is consistent with that indicated by the \ion{Mg}{2} line. They derived a gas mass of 1.5 $\times$ 10$^{10}$ $M_{\odot}$ by adopting the CO excitation model from SDSS  J114816.64$+$525150.3 \citep{Riechers2009} and a conversion factor of 0.8 $M_{\odot} (\rm K\ km\ s^{-1}\ pc^{2})^{-1}$. \citet{Wang2013} marginally resolved this quasar in ALMA [\ion{C}{2}]  observations with resolution of $0\farcs7$. Both the line width and the redshift are consistent with those from the CO (6$-$5) observations. Previous measurements can be seen in Table \ref{tab}. The [\ion{C}{2}]  detection reveals a dynamical mass of 12.5 $\times$ 10$^{10}$ $M_{\odot}$ with an approximately estimate of the inclination angle (56 $\degr$, determined from the ratio between the minor and major axis), suggesting a $M_{\rm BH}/M_{\rm bulge}$ value that is higher than the local value. However, the limit spatial resolution and poor constraint on the inclination angle introduced large uncertainties in the calculation of gas velocity and host galaxy dynamical mass. This is improved by our new ALMA observations presented here.

The outline of this paper is as follows. In Section \ref{sec2}, we present our ALMA Cycle 1 observations, and combine with our ALMA Cycle 0 data \citep{Wang2013} to measure the dust continuum and [\ion{C}{2}]  line emission. In Section \ref{sec3}, we discuss the ISM distribution and investigate gas dynamics by applying a tilted ring model to the [\ion{C}{2}]  velocity map. In Section \ref{sec4}, we summarize our results. Throughout the paper we adopt a $\Lambda$CDM cosmology with $H_{0}$ = 71 km s$^{-1}$ Mpc$^{-1}$, $\Omega_{\rm M}$ = 0.27 and $\Omega_{\Lambda}$ = 0.73 \citep{Spergel2007}.

\section{Observations and results}
\label{sec2}

\subsection{ALMA Observations and Data Reduction}

We imaged the [\ion{C}{2}] line emission ($\nu_{\rm rest}$ = 1900.5369 GHz) of J1319+0950 in August 2014. We used the ALMA band-6 receivers with 34 12 m antennas in the C34-6 configuration. We tuned one of the 2 GHz spectral windows to the redshifted [\ion{C}{2}]  line frequency of $\nu_{\rm obs}$ = 266.443 GHz (we adopted the redshift from \citealt{Wang2013}), and used the other three spectral windows to observe the continuum. The total on-source integration time was 0.6 hours. We calibrated the flux scale based on observations of Titan. The flux calibration uncertainty is $\la$ 15$\%$ for ALMA Cycle 0 J1319$+$0950  [\ion{C}{2}] observations \citep{Wang2013}, and the typical flux calibration uncertainty is better than 10$\%$ for ALMA Cycle 1 observations \citep{Lundgren2012}. For our combined data, we considered a calibration uncertainty $\sim$ 15$\%$. The phase was checked  by observing a nearby phase calibrator, J1347$+$1217. The data were reduced using the Common Astronomy Software Application (CASA\footnote{\url{https://casa.nrao.edu/}}; Version 4.5.0) pipeline.  We subtracted the dust continuum under the [\ion{C}{2}]  line emission in the uv-plane, and binned the data to a channel width of 62.5 MHz ($\sim$ 70 km s$^{-1}$) to optimize the data signal-to-noise ratio (S/N) per velocity bin and the sampling of the FWHM of [\ion{C}{2}] spectrum line. We then combined the new data with ALMA Cycle 0 data \citep{Wang2013}, and made the continuum image and line image data cube from the combined data using the CLEAN task in CASA with robust weighting (robust = 0.5). The synthesized beam size of the final [\ion{C}{2}]  image is $0\farcs28$ $\times$ $0\farcs22$,  corresponding to 1.61 kpc $\times$ 1.27 kpc at the quasar redshift. The 1-$\sigma$ noise is 0.22 mJy beam$^{-1}$ per 62.5 MHz for the line, and 0.03 mJy beam$^{-1}$ for the continuum.

\subsection{Results}

\begin{figure*}
\subfigure{\label{J1319ciiline}\includegraphics[scale=0.32]{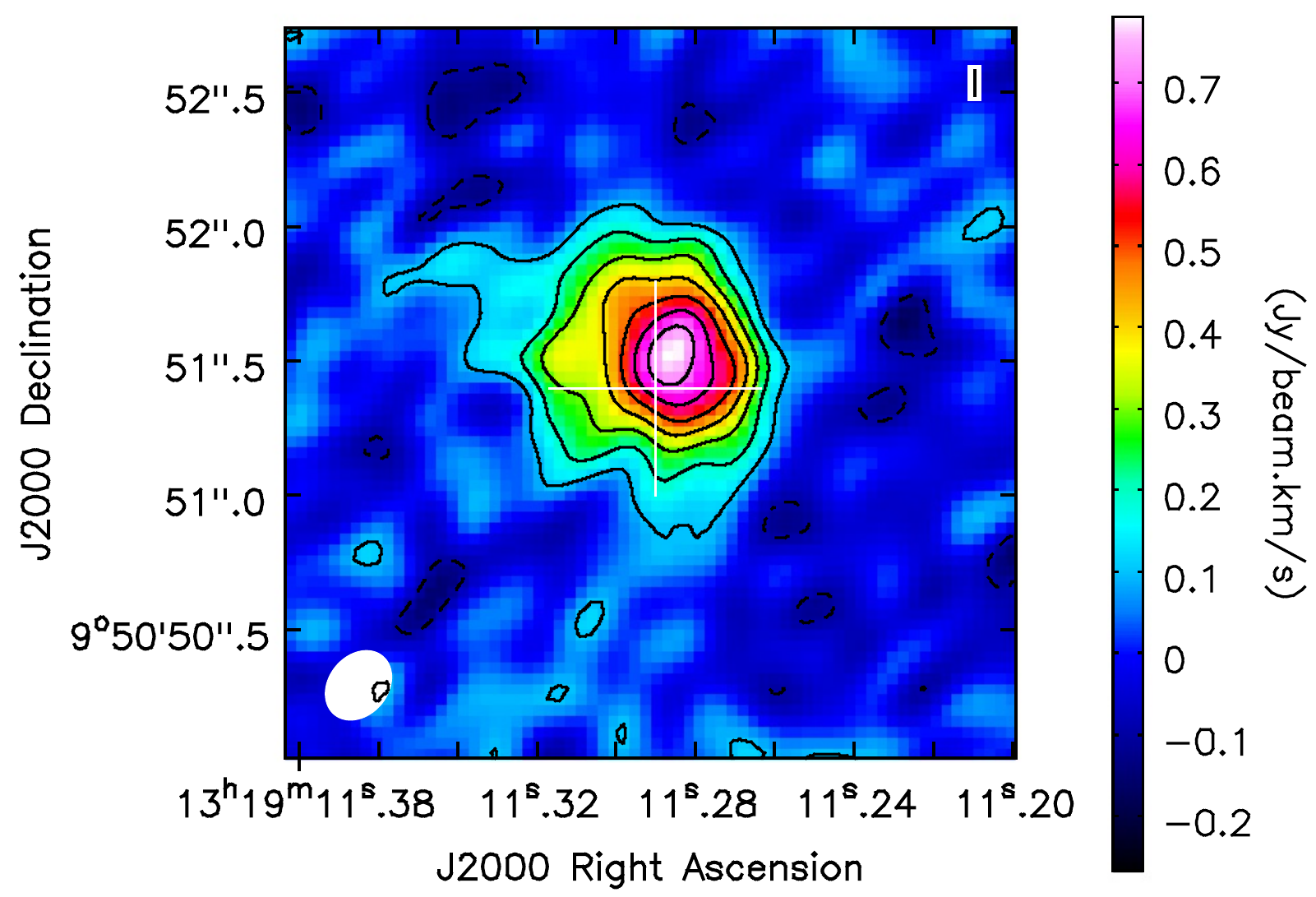}}
\subfigure{\label{J1319con}\includegraphics[scale=0.32]{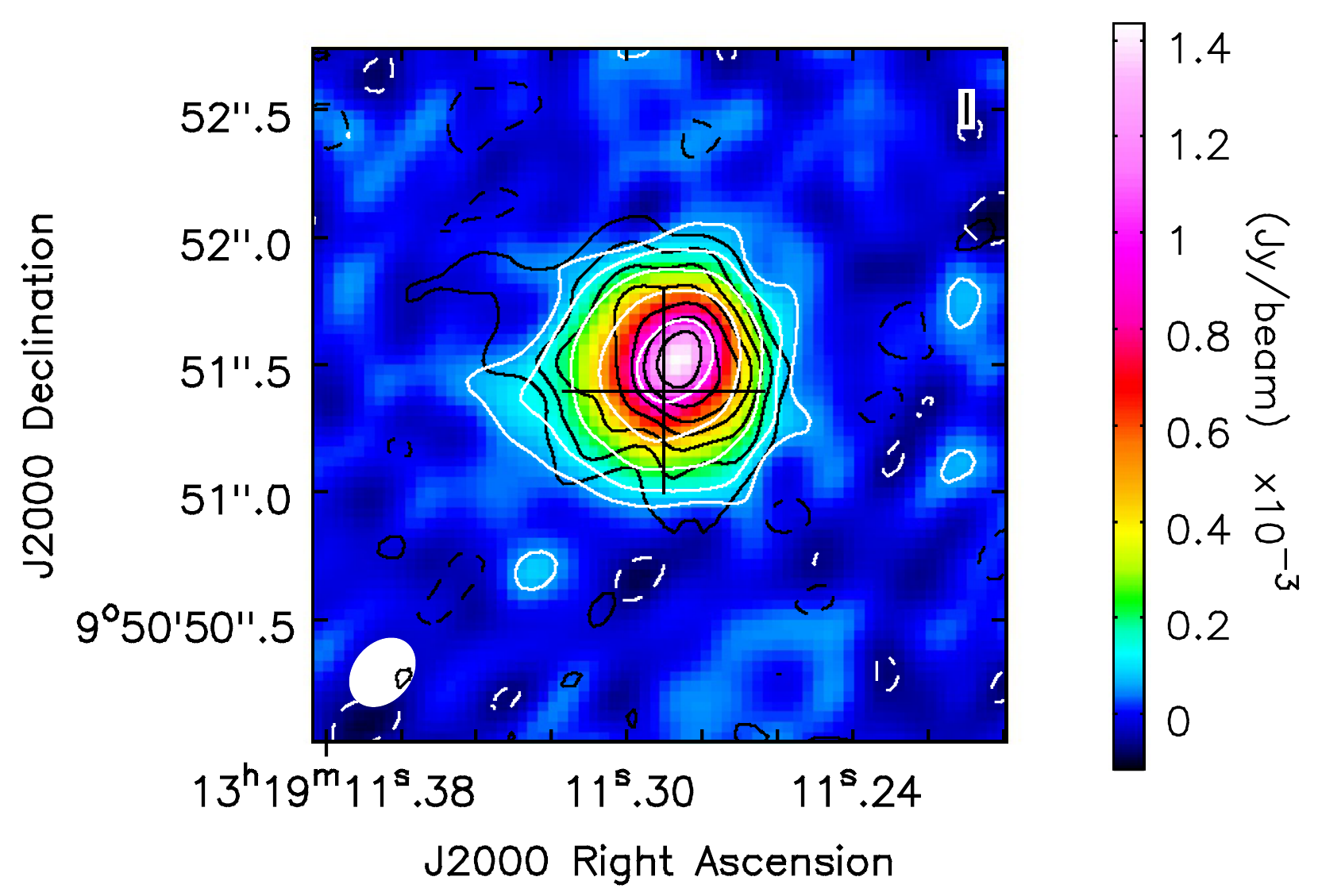}} 
\subfigure{\label{J1319spe}\includegraphics[scale=0.26]{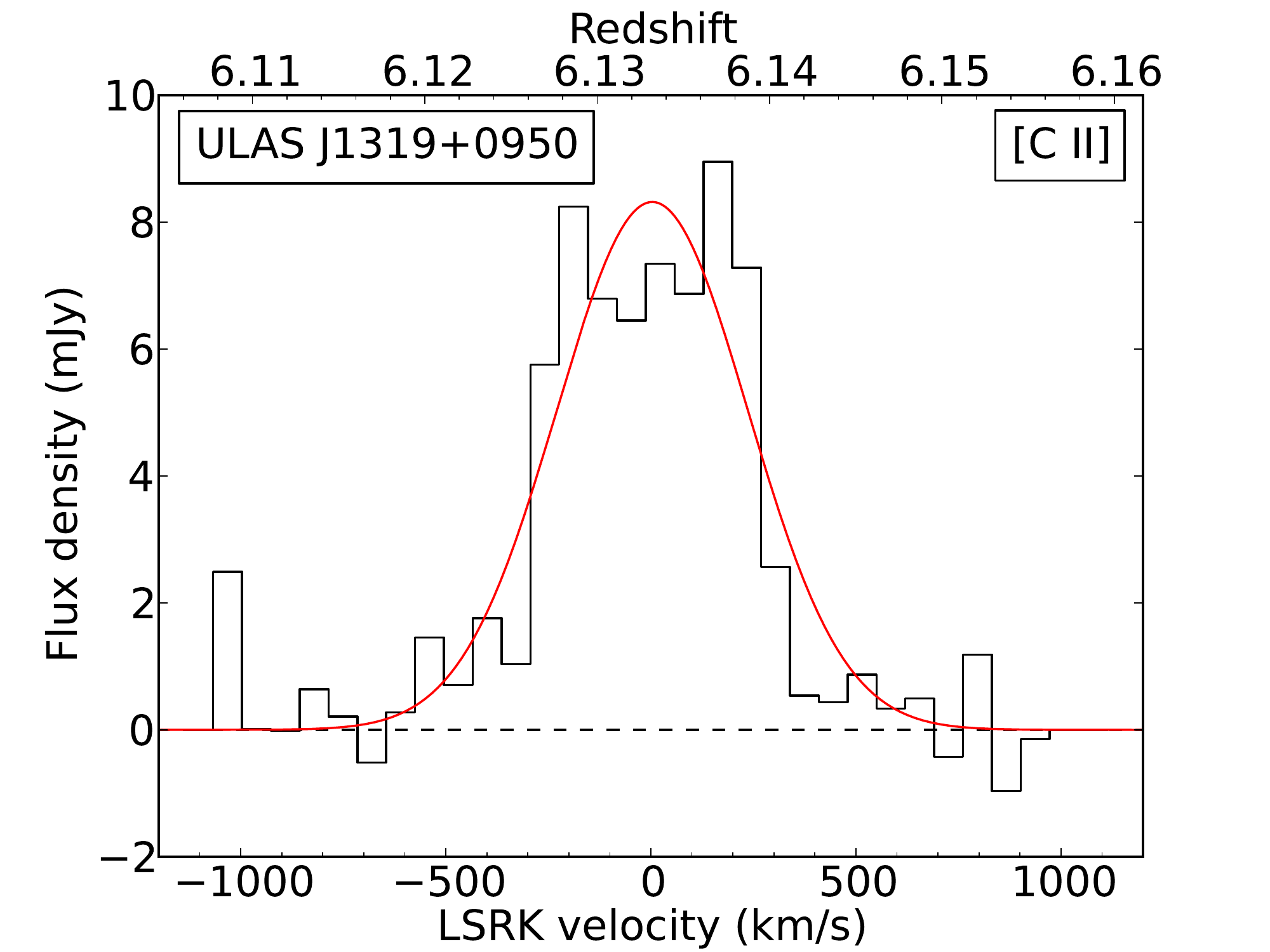}} 
\caption{LEFT: [\ion{C}{2}] velocity-integrated map. The white cross is the infrared position presented by \citet{Mortlock2009}. The bottom left ellipse shows the size of the restoring beam of $0\farcs28$ $\times$ $0\farcs22$. Contour levels are  [$-$2, 2, 4, 6, 8, 10, 12, 14] $\times$ 0.05 Jy beam$^{-1}$ km s$^{-1}$.
	  CENTER: Dust continuum map. The black cross  is the infrared position from \citet{Mortlock2009}. The bottom left ellipse shows the restoring beam size of $0\farcs30$ $\times$ $0\farcs22$. The white contours are [$-$2, 2, 4, 8, 16, 32] $\times$ 30 $\mu$Jy beam$^{-1}$. The over-plotted black contours are the same with those in the left panel. 
	      RIGHT: [\ion{C}{2}] line spectrum (black histogram) over-plotted with the best-fit Gaussian (red line). The LSRK velocity scale is relative to the [\ion{C}{2}] redshift from our ALMA Cycle 0 observations \citet{Wang2013}.}
\label{mapbyalma}
\end{figure*}

The [\ion{C}{2}]  line emission and the dust continuum from the combined data are both spatially resolved. We list the observational results in Table \ref{tab}. The velocity-integrated map of the [\ion{C}{2}]  line emission is presented in the left panel of Figure \ref{mapbyalma}. We fitted the [\ion{C}{2}]  line emission with a 2-D Gaussian,   which yielded a deconvolved source size that is slightly larger than the marginally resolved [\ion{C}{2}]  source size from our ALMA Cycle 0 observations \citep{Wang2013}. 

We integrated the intensity from the [\ion{C}{2}] line image data cube including pixels determined in the line-emitting region above 2-$\sigma$ in the [\ion{C}{2}] velocity-integrated map. The resulting line spectrum is shown as a black histogram in the right panel of Figure \ref{mapbyalma}, with the best-fit Gaussian profile superposed. The Gaussian fit line width is a little larger than, but consistent with our previous  Cycle 0 observations \citep{Wang2013}. The [\ion{C}{2}]  redshift agrees with the result in \citet{Wang2013}. The [\ion{C}{2}]  line flux calculated from the Gaussian fit is consistent with our previous ALMA observations at $0\farcs7$ resolution \citep{Wang2013} within the calibration uncertainty ($\sim$ 15$\%$). We also got a consistent value by calculating the total flux within the 2-$\sigma$ region in the [\ion{C}{2}] intensity map. It is clear that the line profile is flat at the peak in the velocity range from $-$118 km s$^{-1}$ to 93 km s$^{-1}$ (channel centres). A similar [\ion{C}{2}]  line profile was also found in a $z$ = 4.6 quasar \citep{Kimball2015}. Such a profile suggests that the [\ion{C}{2}] line emission originates from a rotating gas disk (see Section \ref{sec3} for a full analysis).

\begin{figure}
\centering
\includegraphics[scale=0.38]{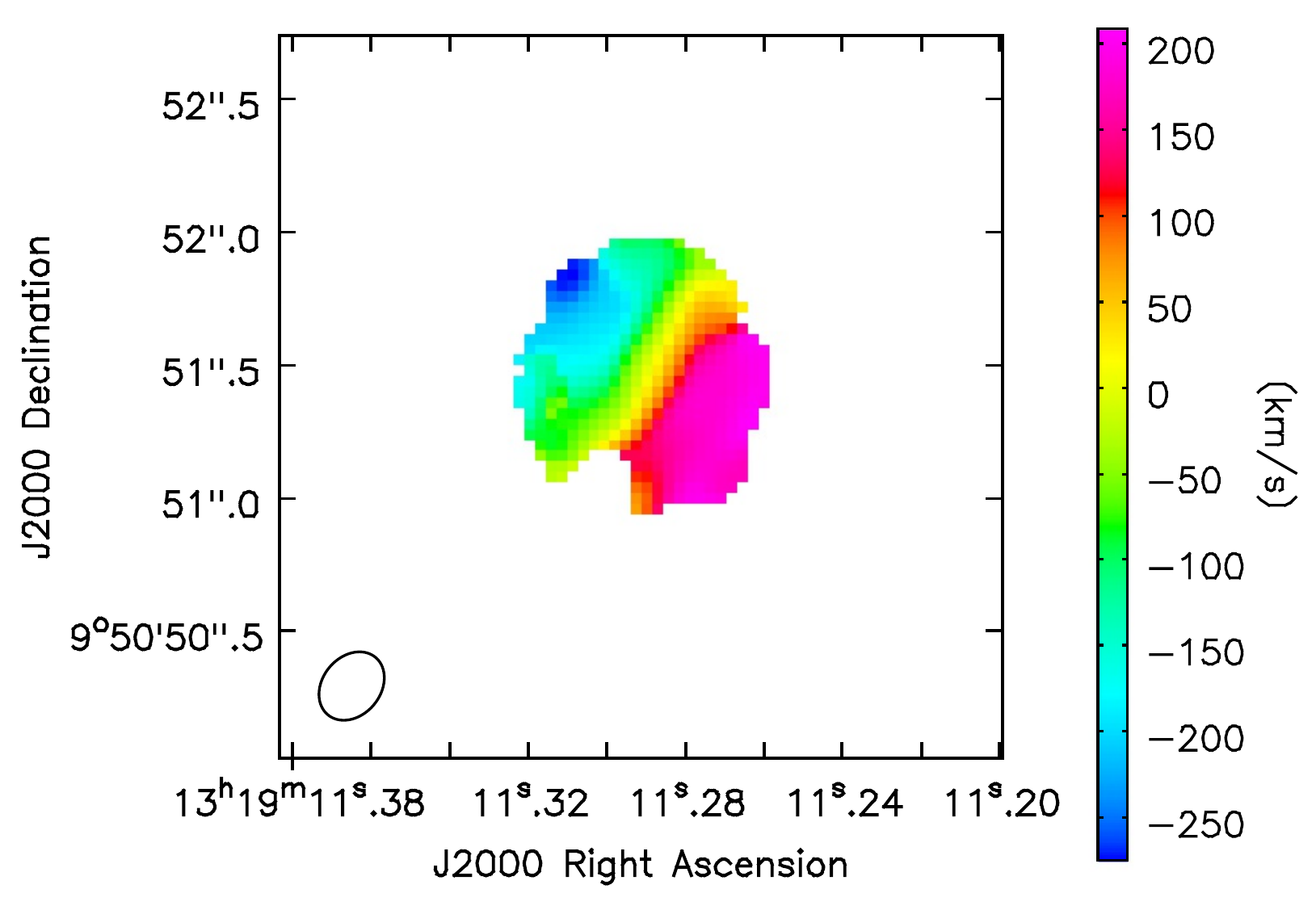}
\caption{Mean gas velocity map based on Gaussian fit.}
\label{mom12}
\end{figure}

\begin{figure*}
\centering
\subfigure{\includegraphics[scale=0.3]{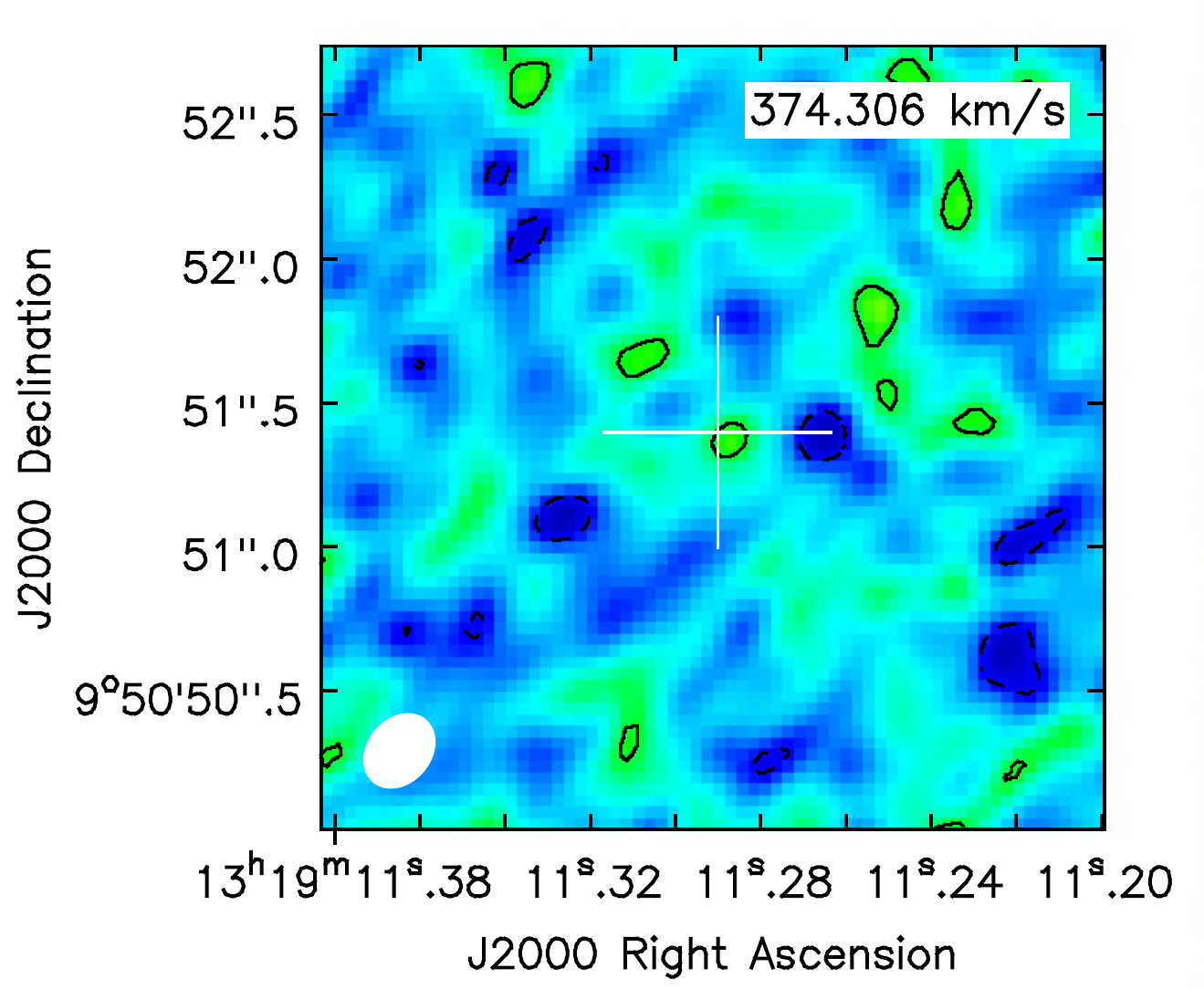}}
\subfigure{\includegraphics[scale=0.3]{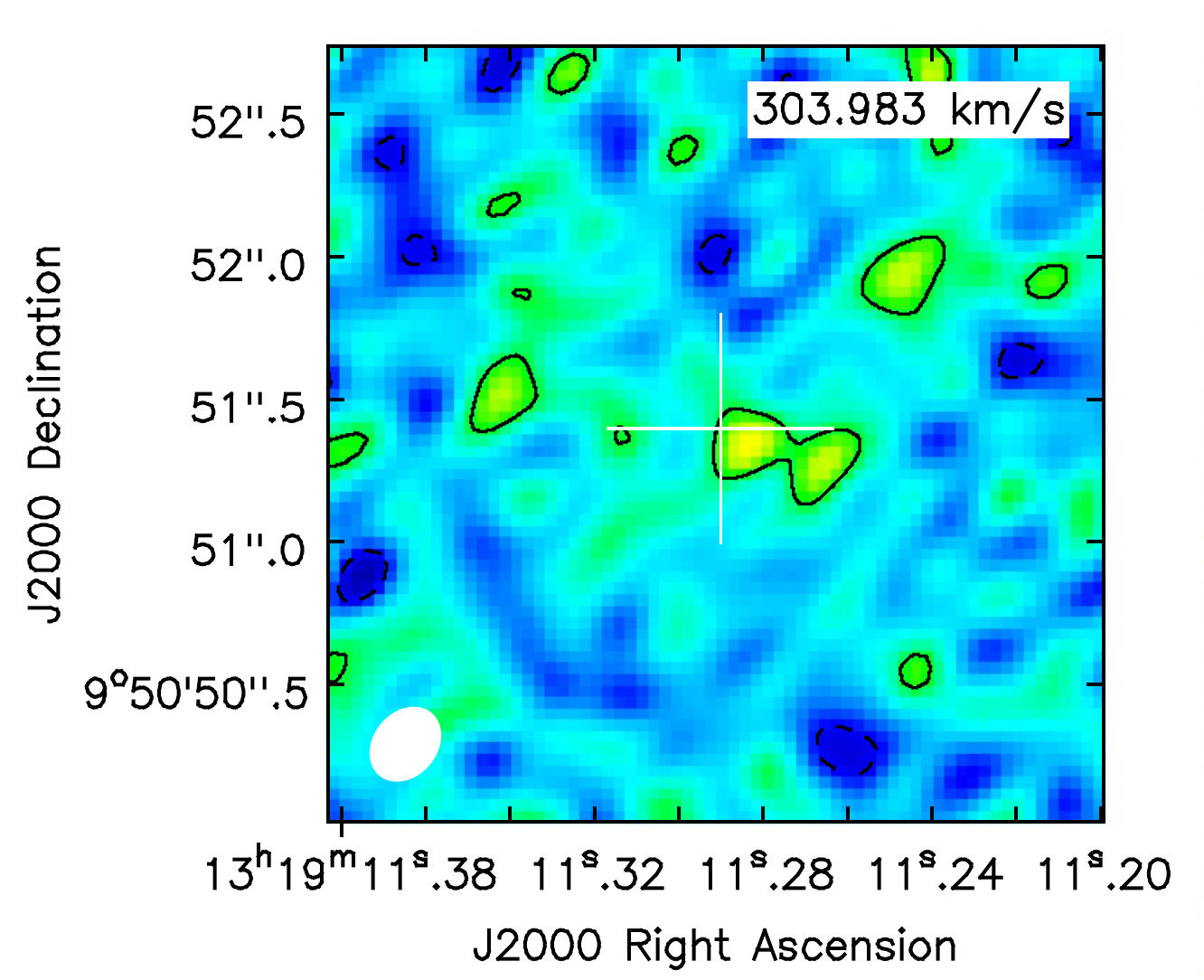}}
\subfigure{\includegraphics[scale=0.3]{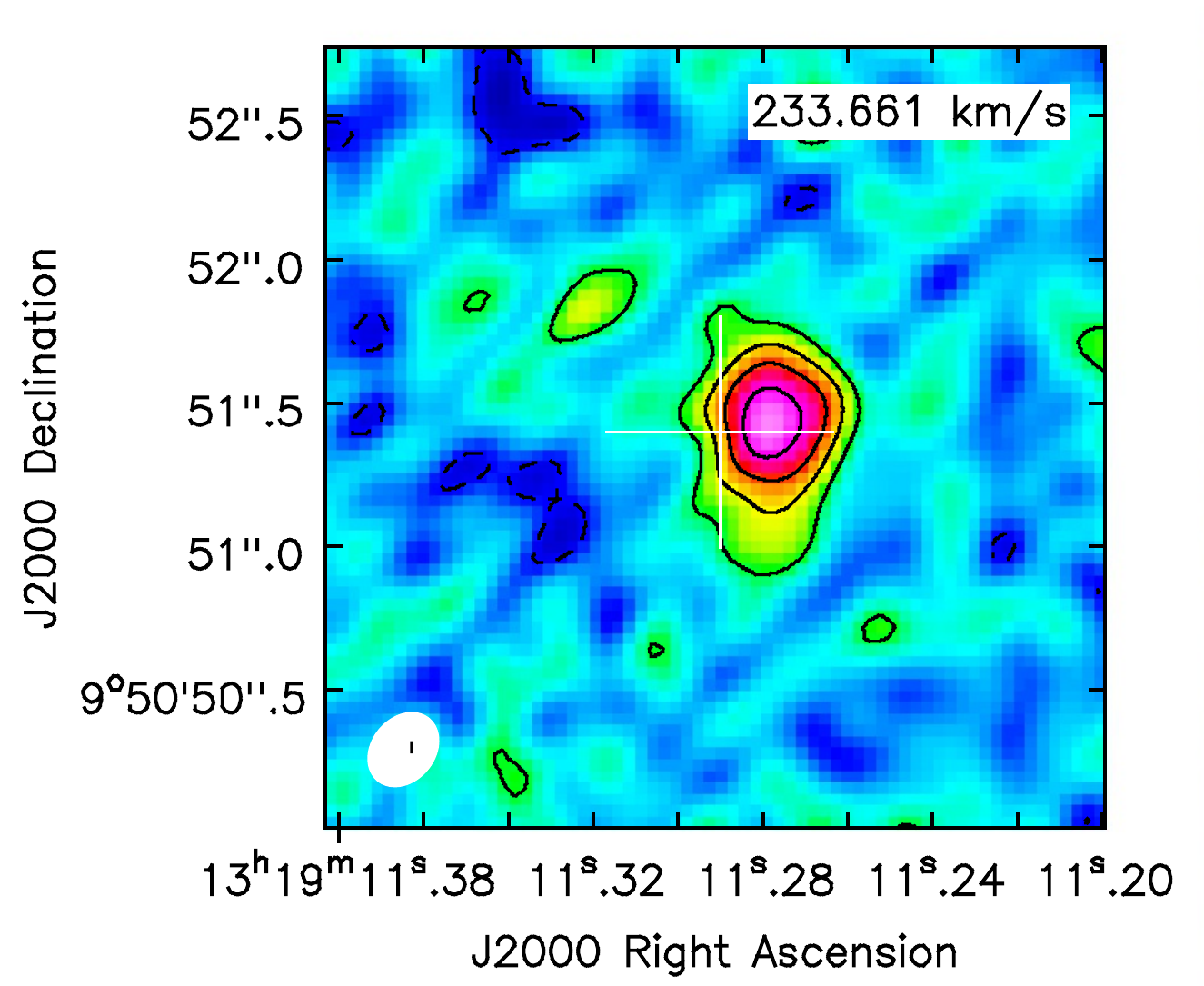}}
\subfigure{\includegraphics[scale=0.3]{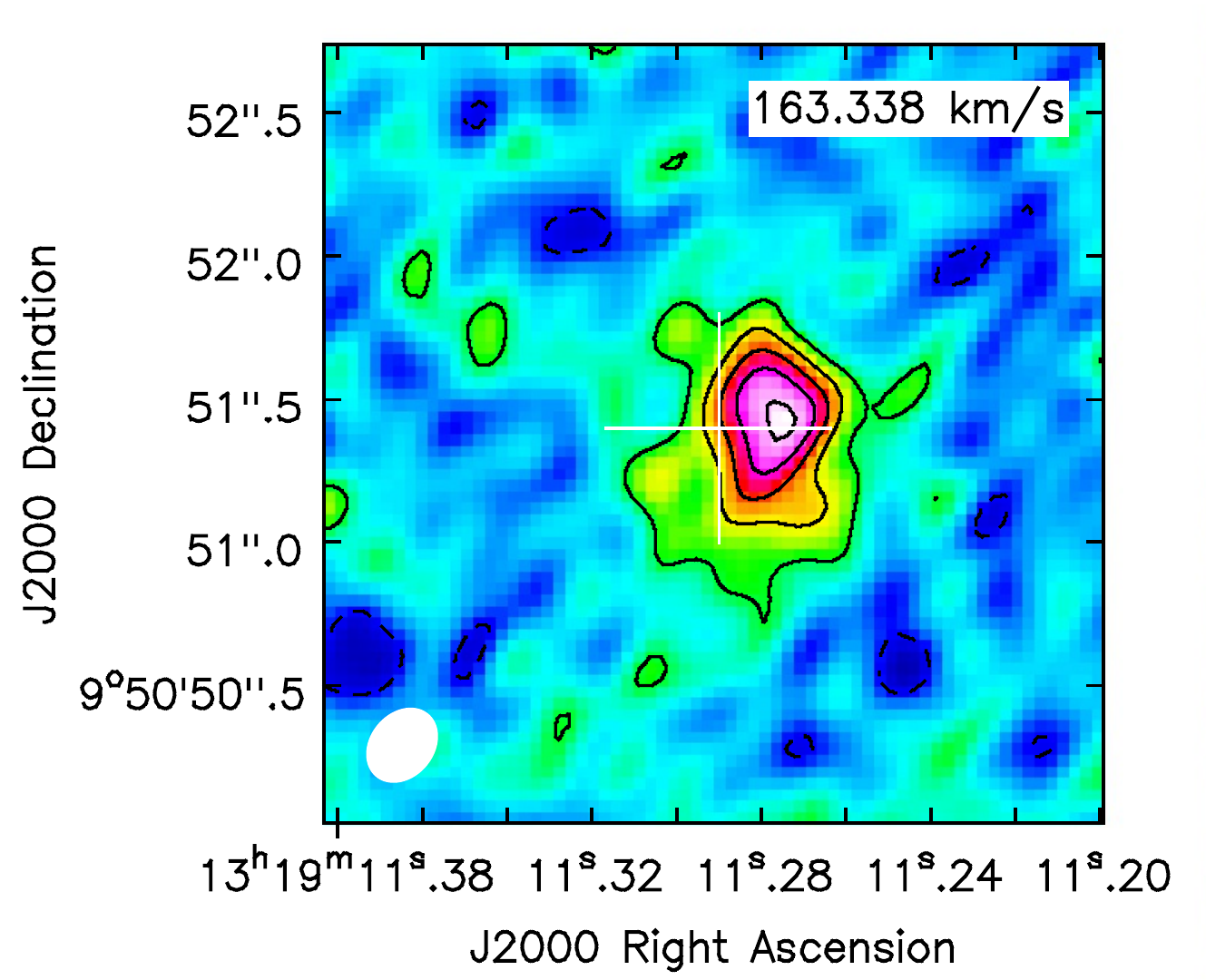}}
\subfigure{\includegraphics[scale=0.3]{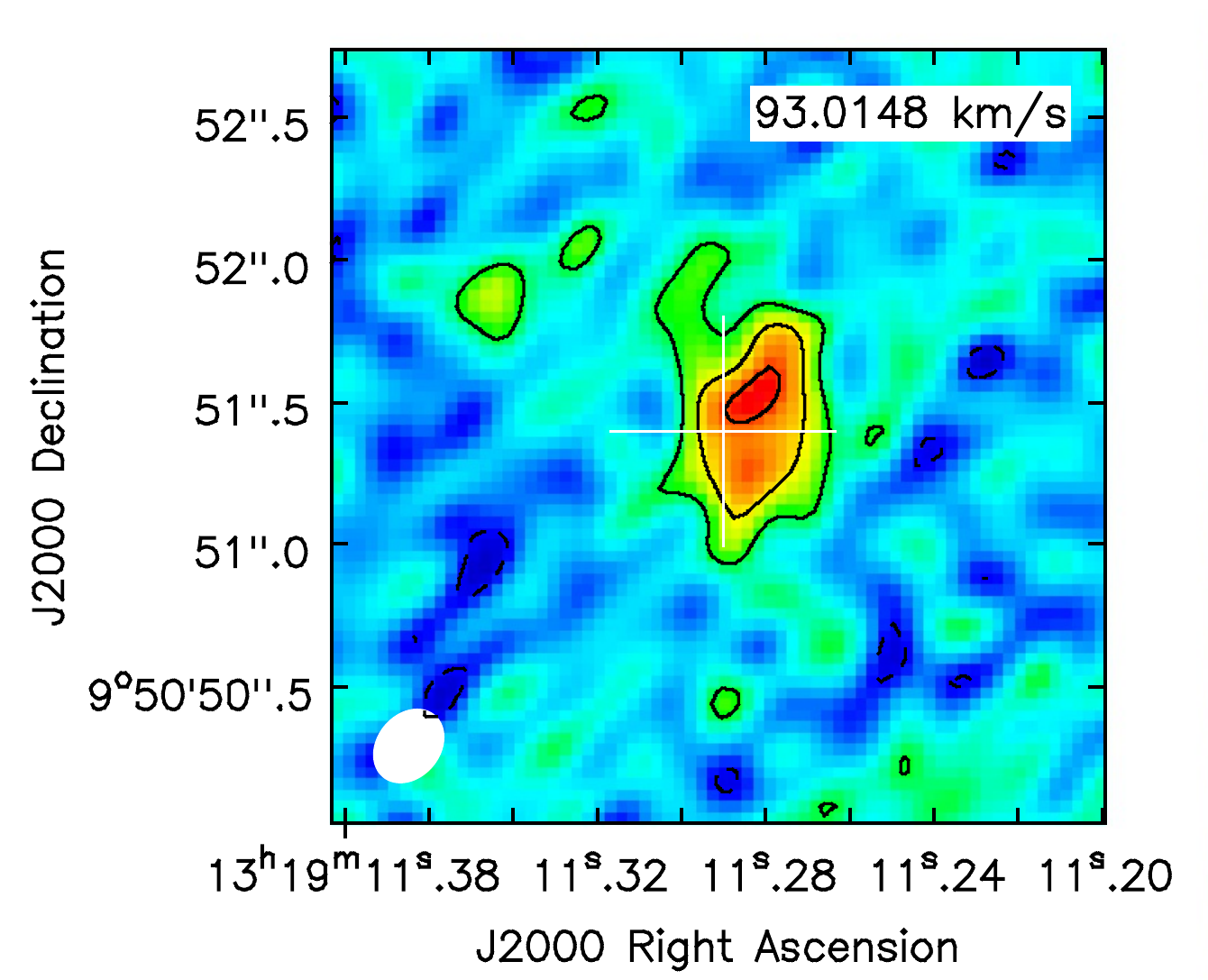}}
\subfigure{\includegraphics[scale=0.3]{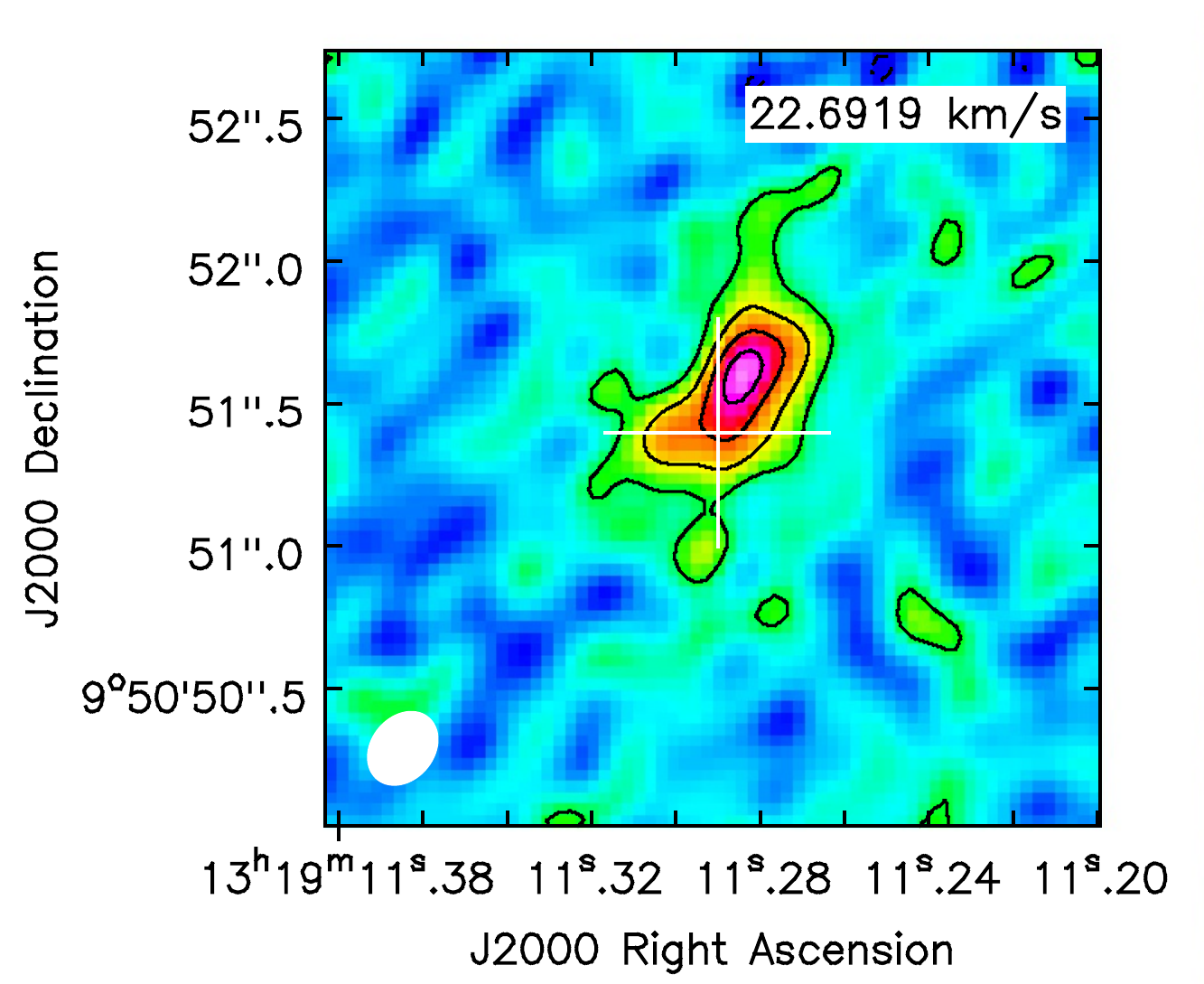}}
\subfigure{\includegraphics[scale=0.3]{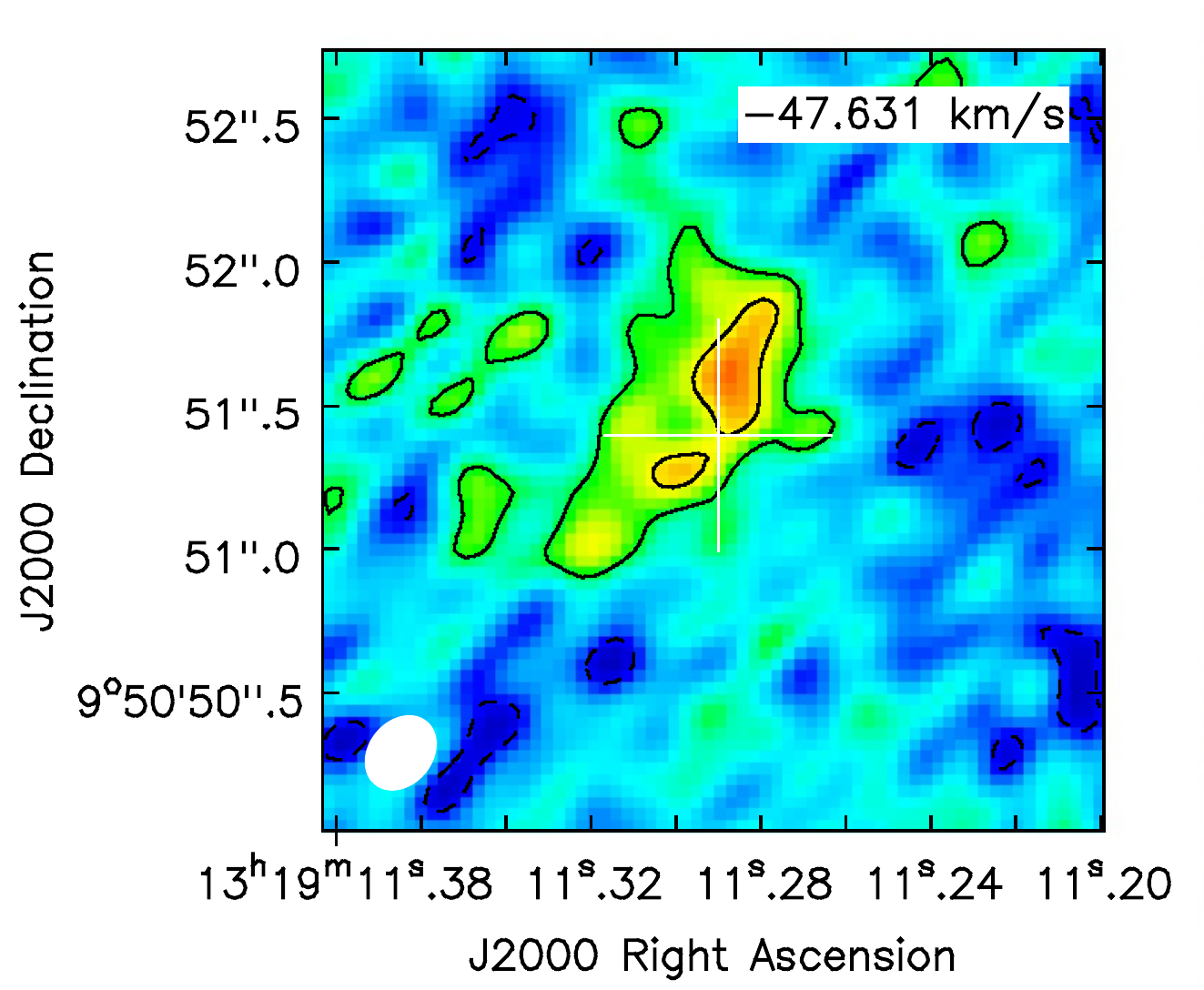}}
\subfigure{\includegraphics[scale=0.3]{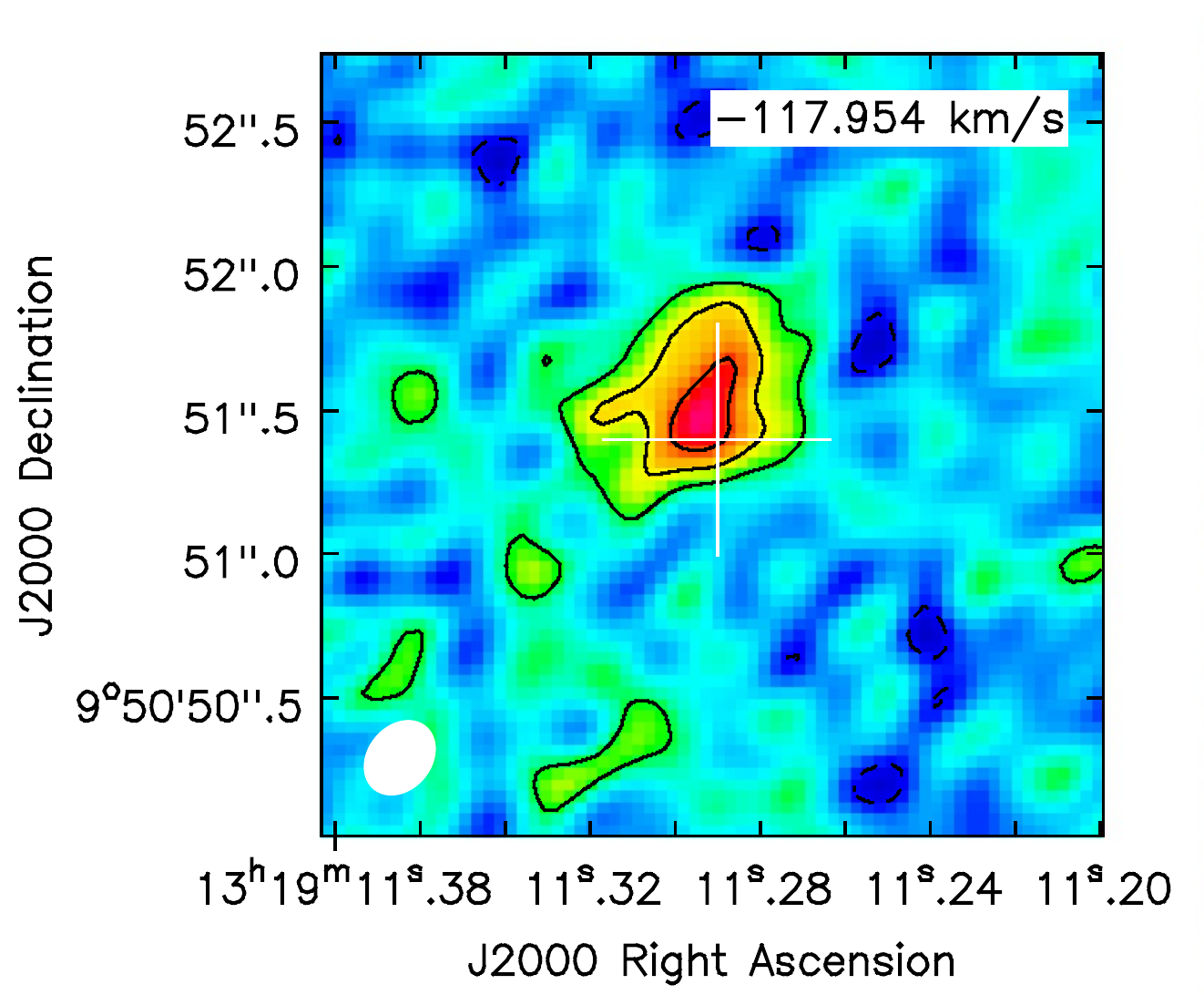}}
\subfigure{\includegraphics[scale=0.3]{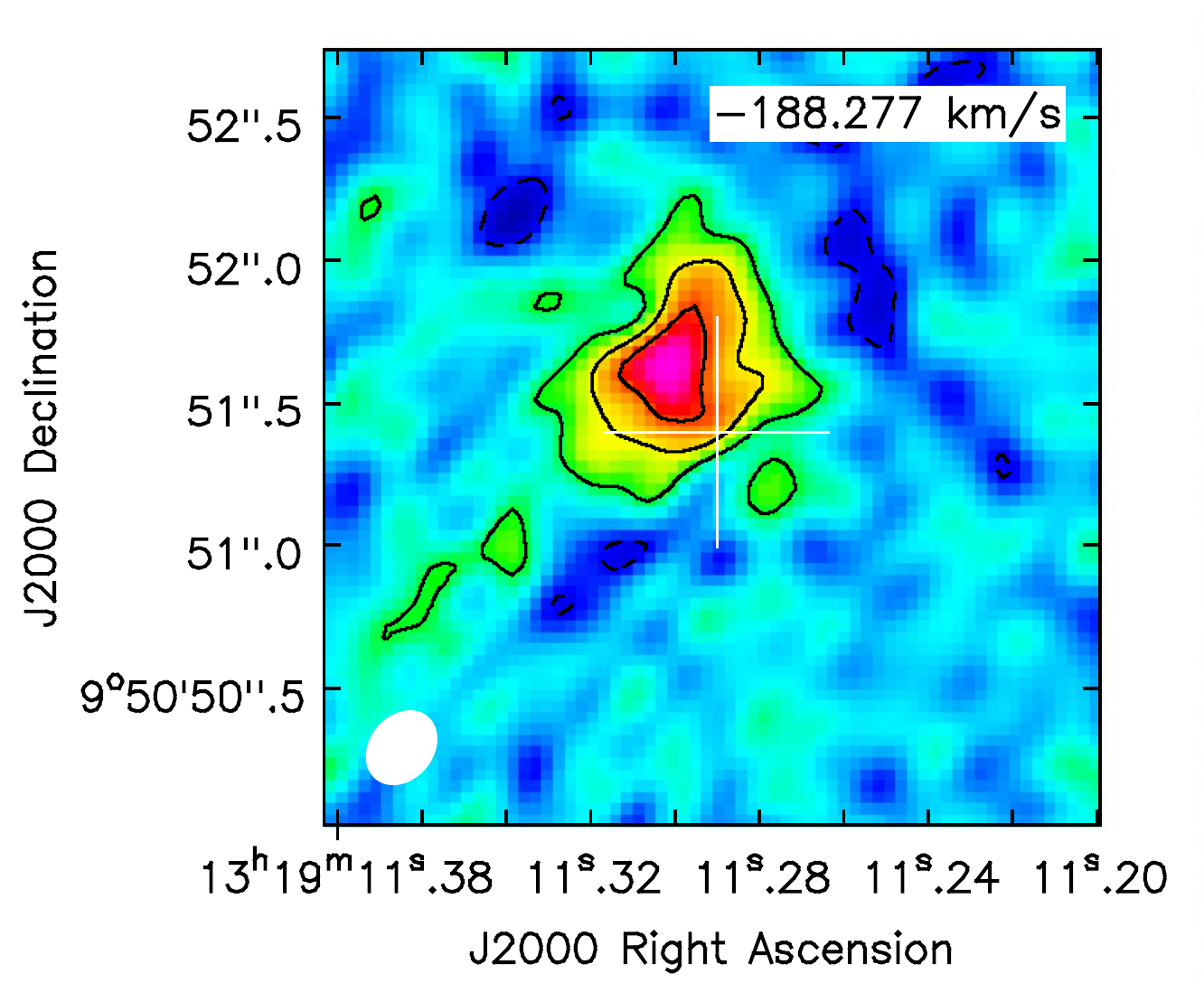}}
\subfigure{\includegraphics[scale=0.3]{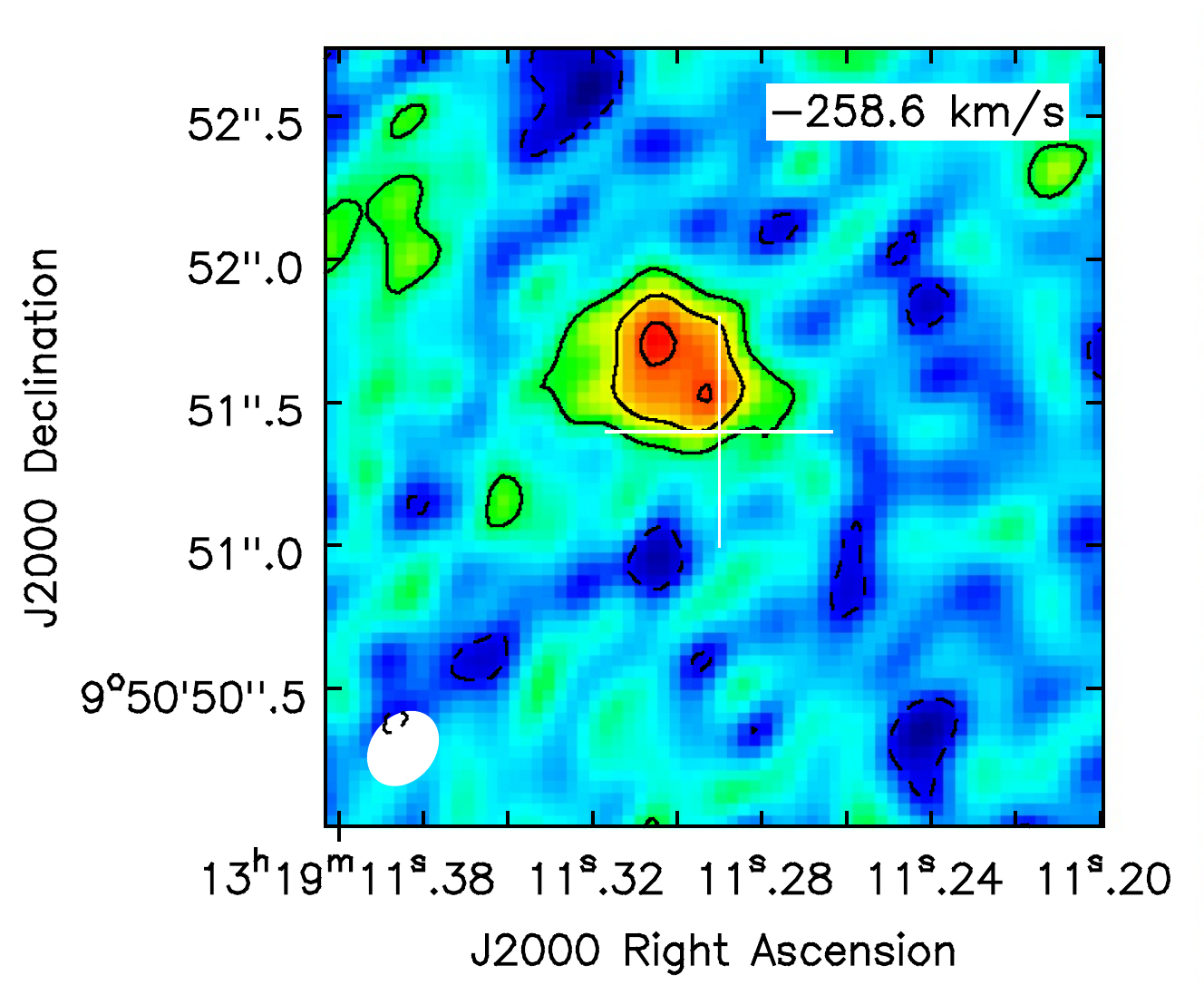}}
\subfigure{\includegraphics[scale=0.3]{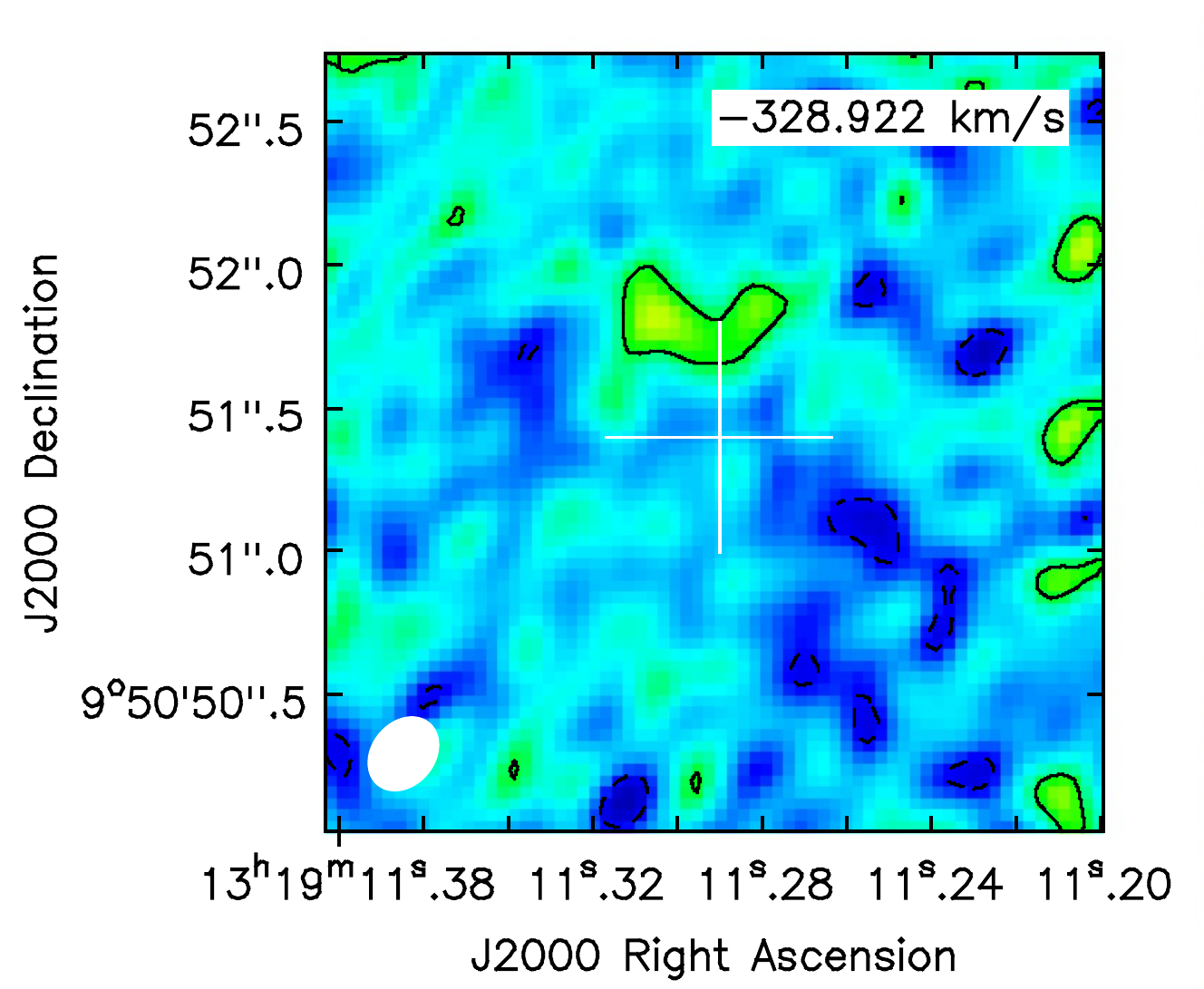}}
\subfigure{\includegraphics[scale=0.3]{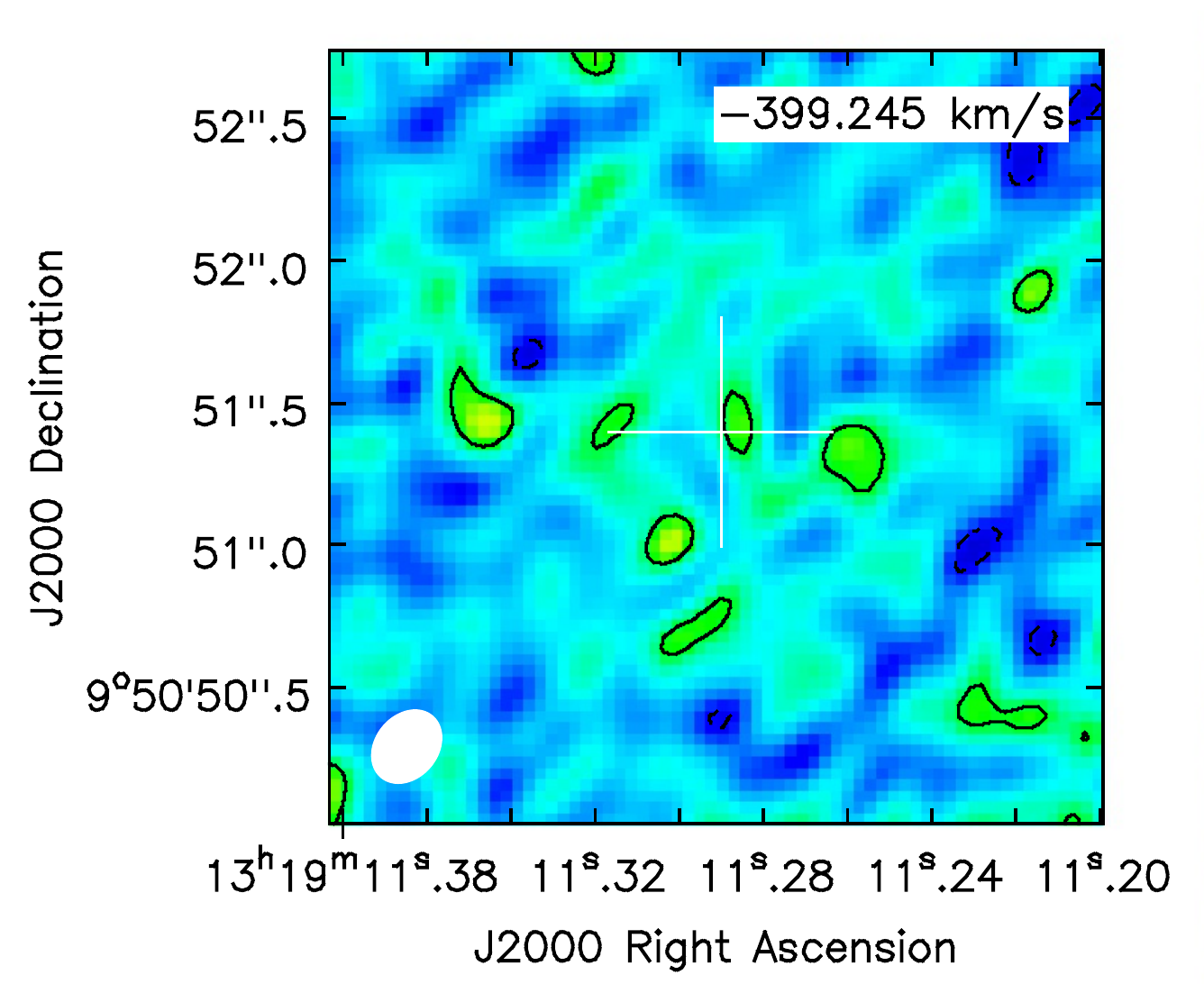}}
\caption{Channel maps of the [\ion{C}{2}]  line emission in the velocity range from 374 km s$^{-1}$ to $-$399 km s$^{-1}$. The velocity takes [\ion{C}{2}] redshift from \citet{Wang2013} as a reference. The channel width is $\sim$ 70 km s$^{-1}$. The contour levels are [$-$2, 2, 4, 6, 8, 10] $\times$ 0.22 mJy beam$^{-1}$. The white cross represents the UKIRT quasar position \citep{Mortlock2009}. The [\ion{C}{2}]  line emission is clearly detected in the central 8 channels, and the emission peak moves from West to East, shifting about $0\farcs4$ from 234 km s$^{-1}$ to $-$259 km s$^{-1}$.}
\label{channelmap}
\end{figure*}

Figure \ref{mom12}  shows the mean gas velocity map with a clear velocity gradient. It was made using the AIPS\footnote{\url{http://www.aips.nrao.edu/}} XGAUS task with 2-$\sigma$ flux cut at each position by Gaussian spectral fit. We also show the [\ion{C}{2}]  line channel maps in Figure \ref{channelmap}. They suggest a clear [\ion{C}{2}]  line emission shift ($\sim$ $0\farcs4$) from 234 km s$^{-1}$ to $-$259 km s$^{-1}$, which is consistent with the velocity map.

We present the dust continuum map in the middle of Figure \ref{mapbyalma}. A 2-D Gaussian fit shows a source size that is a little bigger than, but consistent with the result in the Cycle 0 detection \citep{Wang2013}. The total dust continuum emission is comparable to the emission detected in the previous $0\farcs7$ resolution observations \citep{Wang2013} considering the $\sim$ 15$\%$ calibration uncertainty. We plotted the continuum and [\ion{C}{2}]  contours (white and black lines) over the dust continuum map. The peak of the dust continuum emission is approximately consistent with that of the [\ion{C}{2}] line emission. However, the [\ion{C}{2}]  line emission looks more irregular than the dust continuum even in high S/N regions (e.g., $>$ 4-$\sigma$). This may indicate different distributions between the [\ion{C}{2}]-emitting gas and the dust component in the central few kpc region.

\begin{deluxetable}{lcc}
\tabletypesize{\scriptsize}
\tablewidth{0pc}
\tablecaption{Measured parameters of J1319$+$0950\label{tab}}
\tablecolumns{3}
\tablehead{
\colhead{Parameter} &\multicolumn{2}{c}{Value}
}
\startdata
$^{a}m_{\textsc{1450\AA}}$ (mag)&\multicolumn{2}{c}{19.65}\\
$^{b}S_{\textsc{1.4GHz}}$ ($\mu$Jy)&\multicolumn{2}{c}{64 $\pm$ 17}\\
$^{b}S_{\textsc{250GHz}}$ (mJy)&\multicolumn{2}{c}{4.20 $\pm$ 0.65}\\
$^{a}z_{\textup{\mgii\ }}$&\multicolumn{2}{c}{6.127 $\pm$ 0.004}\\
$^{b}z_{\textsc{CO(6$-$5)}}$&\multicolumn{2}{c}{6.1321 $\pm$ 0.0012}\\
$^{b}$FWHM$_{\textsc{CO(6$-$5)}}$ (km s$^{-1}$)&\multicolumn{2}{c}{537 $\pm$ 123}\\
\hline
&\citet{Wang2013}&this work\\
\hline
$z_{\textsc{\cii\ }}$&6.1330 $\pm$ 0.0007&6.1331 $\pm$ 0.0005\\
FWHM$_{\textsc{\cii\ }}$ (km s$^{-1}$)&515 $\pm$ 81&548 $\pm$ 47\\
$S\Delta \nu_{\textsc{\cii\ }}$ (Jy km s$^{-1}$)&4.34 $\pm$ 0.60&$^{c}$4.85 $\pm$ 0.40 \ \ \ $^{d}$4.31 $\pm$ 0.30\\
$S_{\rm con}$ (mJy)&5.23 $\pm$ 0.10&4.72 $\pm$ 0.17\\
Size$_{\textsc{\cii\ }}$ ($''$)&(0.57 $\pm$ 0.07) $\times$ (0.32 $\pm$ 0.15)&(0.62 $\pm$ 0.06) $\times$ (0.51 $\pm$ 0.05)\\
Size$_{\textsc{\cii\ }}$ (kpc)&--&(3.57 $\pm$ 0.35) $\times$ (2.94 $\pm$ 0.29)\\
Size$_{\rm con}$ ($''$)&(0.39 $\pm$ 0.02) $\times$ (0.34 $\pm$ 0.03)&(0.43 $\pm$ 0.02) $\times$ (0.41 $\pm$ 0.02)\\
Size$_{\rm con}$ (kpc)&--&(2.48 $\pm$ 0.12) $\times$ (2.36 $\pm$ 0.12)\\
\enddata
\tablerefs{$^{a}$\citet{Mortlock2009}; $^{b}$\citet{Wang2011fir}; $^{c}$\cii\ line flux from Gaussian fit to the spectral line; $^{d}$\cii\ line flux from the 2-$\sigma$ region in the velocity-integrated map.}
\tablecomments{The source sizes are all in FWHM. The 15$\%$ calibration uncertainty is not included in the error bar of line/continuum flux.}
\end{deluxetable}

\section{Discussion}
\label{sec3}

\subsection{Gas, Dust and Star Formation Distribution}

\citet{Wang2011fir} presented a gas mass of 1.5 $\times$ 10$^{10}$ $M_{\odot}$ by PdBI CO (6$-$5) observations. Adopting the maximal radius of 3.2 kpc derived in our dynamical fit (Section \ref{gd}) and assuming the same size for the [\ion{C}{2}] and CO(2$-$1) emission, we can derive a gas mass surface density of  466 ($\sim$ $10^{2.67}$) $M_{\odot}$ pc$^{-2}$. This is within the typical range of other star-forming systems at low and high redshifts, e.g., $z$ = 1$-$3.5 submillimeter galaxies (SMGs; $10^{2.30}-10^{4.00}$ $M_{\odot}$ pc$^{-2}$; \citealt{Bouche2007}), $z$ = 1$-$2.3 Bzk-selected galaxies ($10^{1.83}-10^{3.42}$ $M_{\odot}$ pc$^{-2}$; \citealt{Daddi2010}; \citealt{Tacconi2010}), and $z$ = 0 starbursts ($10^{2.25}-10^{4.76}$ $M_{\odot}$ pc$^{-2}$; \citealt{Kennicutt1998sl}).

\citet{Wang2013} estimated the FIR luminosity of (10.7 $\pm$ 1.3) $\times$ 10$^{12}$ $L_{\odot}$ by integrating from 42.5 $\mu$m to 122.5 $\mu$m in the rest frame and assuming a modified black body with a dust temperature of 47 K and an emissivity index of 1.6, which corresponds to a 8$-$1000 $\mu$m luminosity of (15.0 $\pm$ 1.8) $\times$ 10$^{12}$ $L_{\odot}$ (\citealt{Beelen2006}). However, we cannot distinguish the FIR emission contributed by the central AGN and star formation activity. We assume a factor $f_{\rm SF}$ (0 $<$ $f_{\rm SF}$ $<$ 1) which represents the fraction of FIR emission powered by the star formation in the nuclear region. Assuming a Salpeter initial mass function (IMF) and using Equation 4 in \citet{Kennicutt1998}, we can calculate a SFR of (2.6 $\pm$ 0.3) $\times$ $f_{\rm SF}$ $\times$ 10$^{3}$ $M_{\odot}$ yr$^{-1}$. With the largest gas disk radius of 3.2 kpc proposed in Section \ref{gd}, we calculate an average SFR surface density of (81 $\pm$ 9) $\times$ $f_{\rm SF}$ $M_{\odot}$ yr$^{-1}$ kpc$^{-2}$. The values of the SFR and SFR surface density could be lower by a factor of 1.7 if we assume a Chabrier IMF \citep{Chabrier2003}. Our source has a very high SFR surface density that can be comparable to the highest values found in samples of SMGs with similar gas mass surface density (\citealt{Bouche2007}; \citealt{Hodge2015}; \citealt{Bothwell2010}), if we assume that all the dust continuum is produced by star formation.

\subsection{Gas Dynamics in the Quasar Host Galaxy}
\label{gd}

\subsubsection{GIPSY modeling of gas dynamics}
\label{gipsy model}

Both the flat-peak line profile in the right panel of Figure \ref{mapbyalma} and the velocity gradient in  Figure \ref{mom12} are consistent with a rotating gas disk. There are also tentative non-rotating structures, e.g., the tail structures in the fifth to seventh channel images in Figure \ref{channelmap}. Deeper imaging of these low surface brightness components will determine if there are indeed non-rotating/tidal-like structures in this system and address if there is evidence of a recent galaxy merger.

In our work, we simply assume that the gas has a pure circular rotation in a gas disk, and fit the velocity field with a tilted ring model \citep{Rogstad1974}. The tilted ring model decomposes a galaxy into many  thin rings, and the dynamic property of each ring at different radii can be described by seven parameters:

$\bullet$ ($x_{\rm 0}, y_{\rm 0}$): the sky coordinates of the rotation centre of the galaxy. 

$\bullet$ $V_{\rm sys}$: the velocity of the centre of the galaxy with respect to the Sun.

$\bullet$ $V_{\rm c}(R)$: the circular velocity at distance $R$ from the centre.

$\bullet$ $\phi(R)$: the position angle of the major axis on the receding half of the galaxy, taken anti-clockwise from the north direction on the sky.

$\bullet$ $i(R)$: the inclination angle between the normal to the plane of the galaxy and the line-of-sight.

$\bullet$ $\theta(R)$: the azimuthal angle related to $i(R),\ \phi(R),\ (x_{\rm 0}, y_{\rm 0})$.

The line of sight velocity $V_{\rm los} (x, y)$ that we observed is a projected value. It is related to the above parameters: 

\begin{equation}
\label{eq1}
V_{\rm los}(x, y) = V_{\rm sys} + V_{\rm c}(R) \sin i \cos\theta
\end{equation}

\begin{equation}
\label{eq2}
\cos\theta = \frac{-(x - x_{\rm 0}) \sin{\phi} + (y - y_{\rm 0}) \cos\phi}{R}
\end{equation}

\begin{equation}
\label{eq3}
R^{2} = (x - x_{\rm 0})^{2} + (y - y_{\rm 0})^{2}
\end{equation}

\begin{figure*}
\centering
\subfigure{\label{a}\includegraphics[scale=0.3]{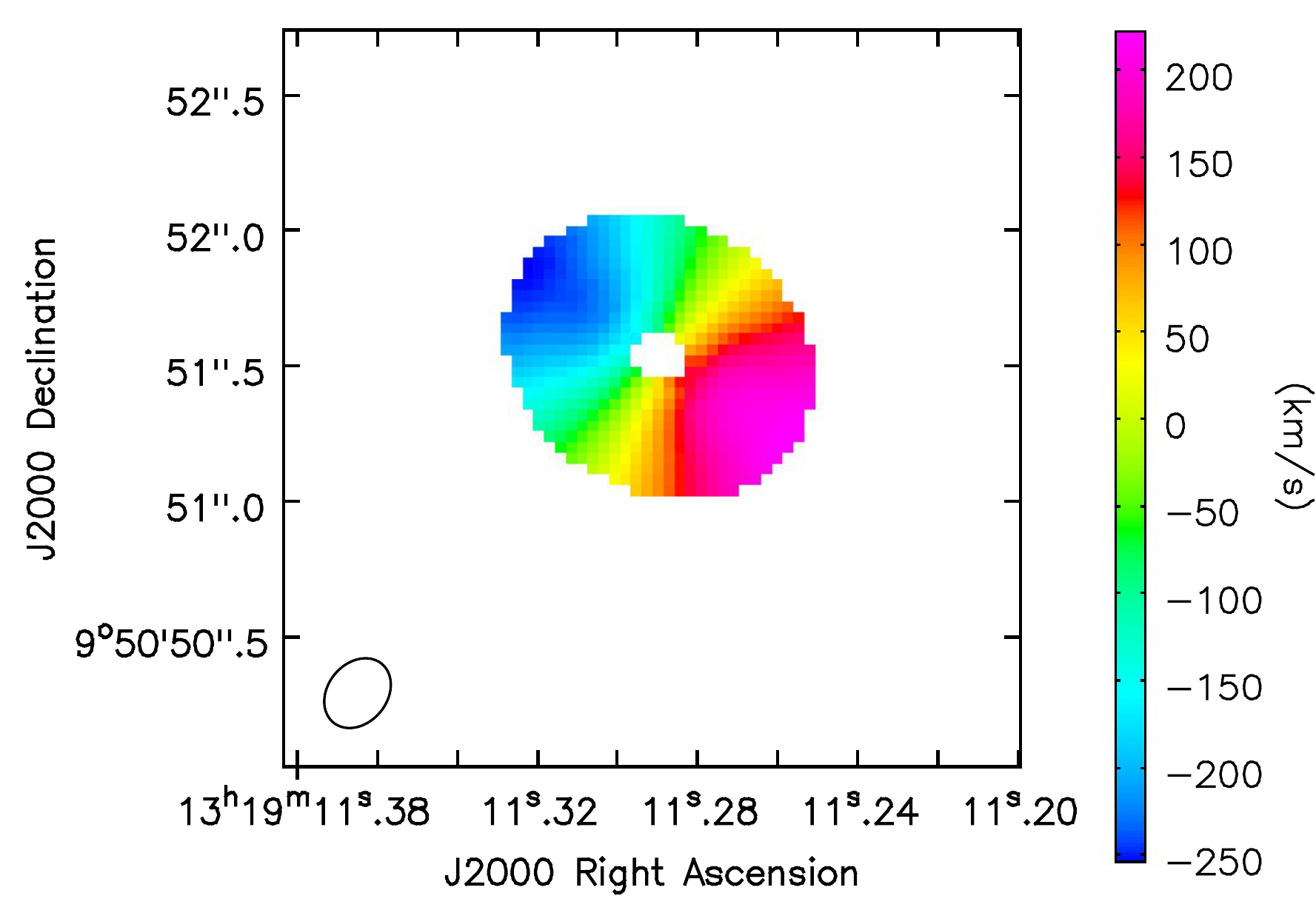}}
\subfigure{\label{b}\includegraphics[scale=0.3]{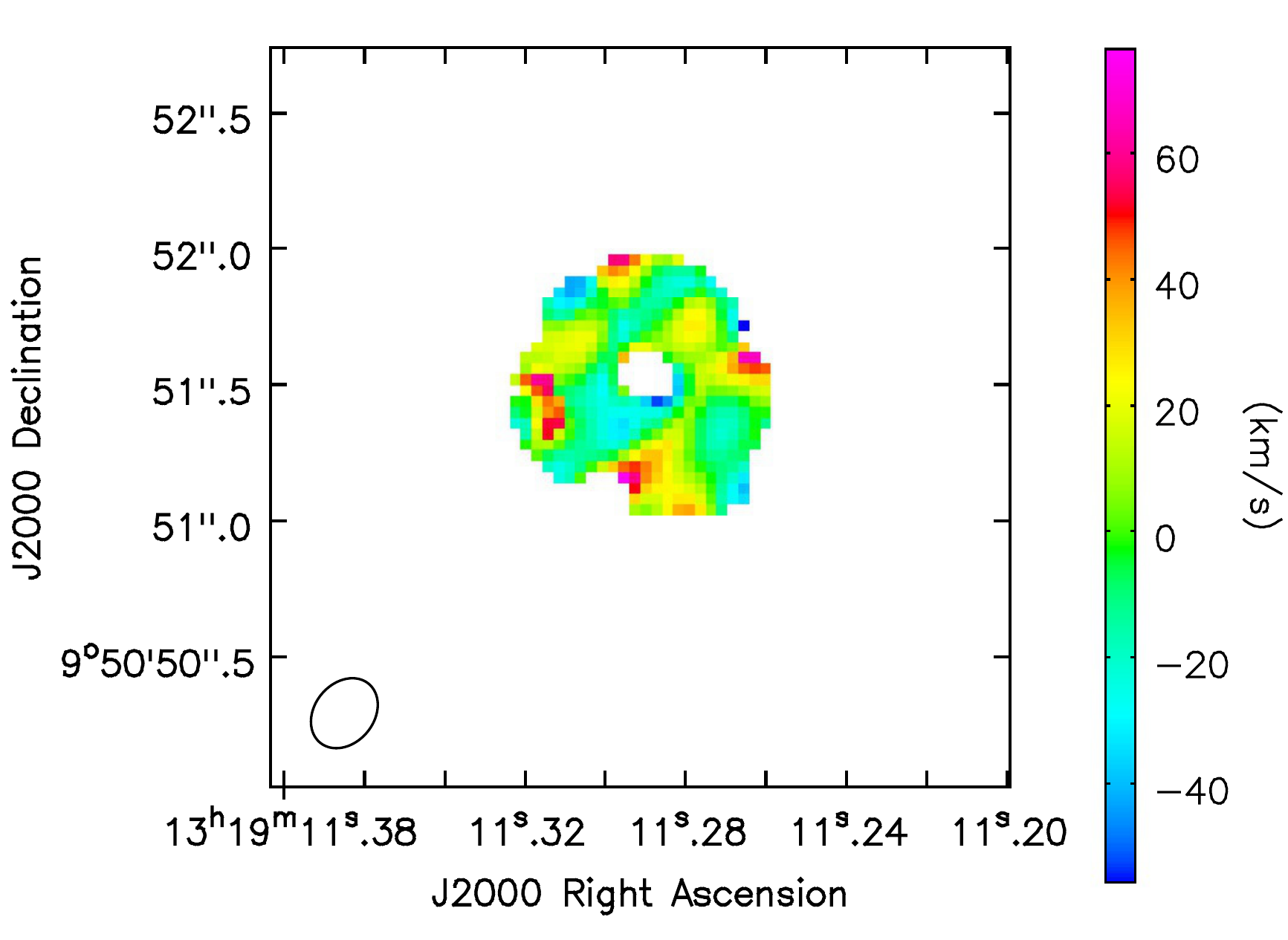}}
\subfigure{\label{c}\includegraphics[scale=0.27]{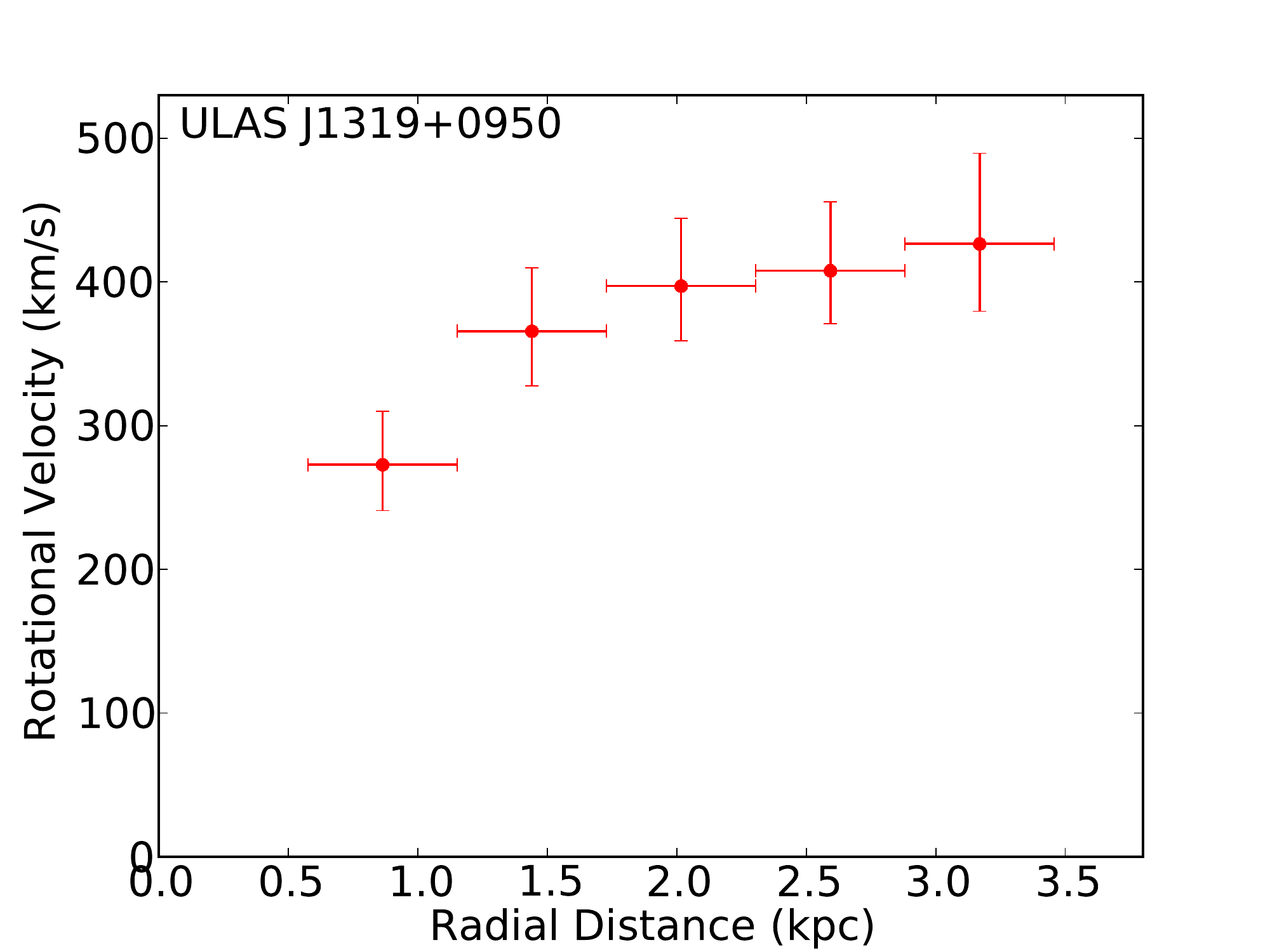}}
\caption{GIPSY modeling result. Panels from left to right: GIPSY modeled velocity map, residual map and rotation curve. In the left and middle panels, the plotted restored beam size is $0\farcs28$ $\times$ $0\farcs22$, the same size as the observed [\ion{C}{2}]  map. There is a hole in the centre of the modeled velocity map, because we do not have enough data in the central region to model the dynamical motion.}
\label{model}
\end{figure*}

We use ROTCUR task in the Groningen Image Processing System (GIPSY\footnote{\url{https://www.astro.rug.nl/~gipsy/}}; \citealt{vanderHulst1992}) to apply the tilted ring model to the observed velocity field to calculate the kinematic parameters. We assume that all rings share the same ($x_{\rm 0}, y_{\rm 0}$), $V_{\rm sys}$, $\phi$, and $i$. We solve for $V_{c}(R)$ in five concentric rings, each with a width of $0\farcs1$ and central radius from $0\farcs15$ to $0\farcs55$. We determine the initial values of $i$, $\phi$, and ($x_{\rm 0}, y_{\rm 0}$) based on the 2-D Gaussian fit to the [\ion{C}{2}] intensity map (left panel of Figure \ref{mapbyalma}), and set the initial value of $V_{\rm sys}$ from the Gaussian fit to the [\ion{C}{2}]  spectrum (right panel of Figure \ref{mapbyalma}). We solve for the five parameters as follows: because $\phi$ and $i$ are correlated, we first  simultaneously determine them by fixing initial values of ($x_{\rm 0}, y_{\rm 0}$) and $V_{\rm sys}$. The final values of $\phi$ and $i$ are calculated as the weighted mean of each $\phi(R)$ and $i(R)$, and the uncertainties are taken as the weighted standard deviations ($\sigma_{sdv}$) of the fitting parameters (we take 1/$\sigma$ as the weighting). Note that only rings with fitting parameter values above 3-$\sigma$ are considered as a successful fit, and are used in the $\phi$ and $i$ calculation. In particular, only two rings are successful for $i$ calculation. The successful $i(R)$ solutions of the two rings are 38 $\pm$ 10 $\degr$ and 32 $\pm$ 6 $\degr$. Since ($x_{\rm 0}, y_{\rm 0}$) and $V_{\rm sys}$ are coupled, we then determine the two parameters simultaneously by fixing $\phi$ and $i$ as the values derived from the previous step. We calculate their final values and uncertainties with the same method above. The quoted errors of these parameters are only fitting-type errors, which do not account for the covariance between these parameters. Similar dynamical analysis with ROTCUR can be seen in Jones et al. (in preparation).

The final fitting values and weighted standard deviations of $V_{\rm sys}$, $\phi$, and $i$ are $-$15 $\pm$ 3 km s$^{-1}$, 237 $\pm$ 4 $\degr$, and 34 $\pm$ 4 $\degr$, respectively. There are other two input values to be declared in ROTCUR: free angle and weighting.  Following the recommendation by \citet{Lucero2015}, we adopt a UNIFORM weighting and an exclusion angle of 0 $\degr$ to use all data with same weighting. 

\subsubsection{Rotation curve}
We obtain the rotation curve adopting the final values of ($x_{\rm 0}, y_{\rm 0}$), $V_{\rm sys}$, $\phi$, and $i$  with ROTCUR. We estimate the error bars of the rotational velocities as follows: first we run ROTCUR adopting our standard values of ($x_{\rm 0}, y_{\rm 0}$), $V_{\rm sys}$ and $\phi$, but change $i$ by $\pm$ 1-$\sigma_{sdv}$. Then we determine the error bars by subtracting these two rotation curves from the one with the best-fit $i$. In addition, we also add the fitting-errors to the final errors. We present the rotation curve in the right panel of Figure \ref{model}. The curve rises to 2 kpc, and then flatten on larger scales. The circular velocity at the largest radius (i.e., 3.2 kpc) is 427 $\pm$ 55 km s$^{-1}$. The left panel of Figure \ref{model} shows our modeled velocity field produced by VELFI task in GIPSY. The residual map is shown in the middle panel with velocity difference less than 30 km s$^{-1}$ across the entire velocity field.

However, we need to point out that the inclination angle ($i$) is calculated as the weighted mean of only two successful $i(R)$ values. Thus, the real uncertainty in $i$ could be much larger than the error bar mentioned above. There should also exist convariance with other parameters as we cannot fit all the parameters independently at the same time. These will result in large uncertainties in the rotational velocities, which are not included in the error bars shown in Figure \ref{model}. A more definite estimate of the error bar of the inclination is undergoing based on model data analysis (Jones, Shao, et al. in preparation). In order to give a more realistic estimate of the uncertainty in the rotational velocity, we check the rotation curve fit with inclination angle values in the range of 26 $\degree$ to 48 $\degree$ which covers the $i(R)$ values and 1-$\sigma$ uncertainties we found with the two successful rings (see Section 3.2.1). The rotational velocity at the largest radius increases to 537 km s$^{-1}$ with $i=26\degree$ and decreases to 331 km s$^{-1}$ with $i=48\degree$.


In addition, the tilted ring model we adopted in this work does not take into account the effect of the synthesized beam. The beam smearing effect could smooth out any rapid change in the velocity field within the beam (\citealt{Bosma1978}; \citealt{Begeman1987}). As was discussed in the extensive studies of \ion{H}{1}-based rotation curves of galaxies, this could affect the inner part of the derived rotation curve, resulting in a shallower slope compared to the intrinsic one (\citealt{Swaters2000}; \citealt{deBlok1997}) and introduce additional uncertainties in the fitting parameters (e.g., inclination angle, rotation velocity, etc) of the inner rings (\citealt{Swaters2009}; \citealt{Kamphuis2015}). However, the beam smearing effect may not play an important role in our measurements of the outer/flat part of the rotation curve, unless the intrinsic rotation curve is non-flat at large radius (e.g., a solid-body type rotation curve found in dwarf galaxies; \citealt{deBlok1997}).

\subsubsection{$M_{\rm BH}$-$M_{\rm dyn}$ relation}

Adopting the rotational velocity obtained with the best-fit $i$ of 34 $\degree$, we calculate the host galaxy dynamical mass within the central 3.2 kpc radius to be $M_{\rm dyn}=13.4  \times 10^{10}\,M_{\odot}$. The dynamical mass is a little bit higher than that estimated by \citet{Wang2013}.  Resolving the gas disk with high resolution imaging is very important to accurately measure the dynamical mass of the distant quasar hosts. 

\begin{figure}
\centering
\includegraphics[scale=0.38]{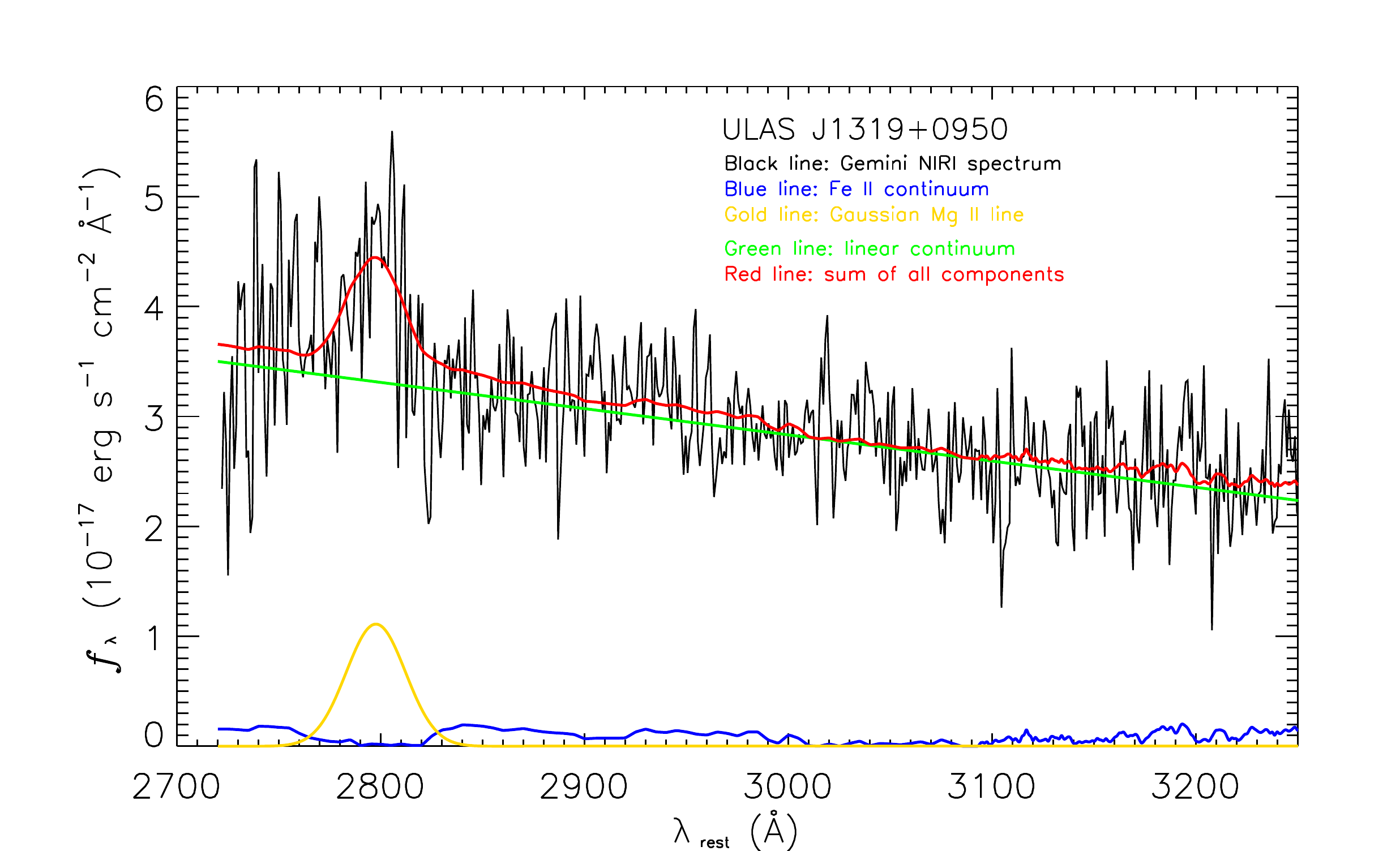}
\caption{Gemini NIRI spectrum of ULAS J1319+0950 (black line) fitted with a linear continuum (green line), a Gaussian \ion{Mg}{2} line (gold line), and an ultraviolet \ion{Fe}{2} template (\citealt{Shen2012}; blue line). The sum of these components can be seen in red line.}
\label{ir}
\end{figure}

To calculate $M_{\rm BH}$, we fit the Gemini NIRI spectrum of ULAS J1319+0950 \citep{Mortlock2009} with a linear continuum,  a Gaussian for \ion{Mg}{2} line, and an ultraviolet \ion{Fe}{2} template based on \citet{Shen2012} (see in Figure \ref{ir}). We derive a width of \ion{Mg}{2} line to be 34.3 $\pm$ 1.6 $\rm \AA$ and $L_{3000 \rm \AA}$ to be (3.8 $\pm$ 1.0) $\times$ 10$^{46}$ erg/s. The final $M_{\rm BH}$ calculated from Equation 3 in \citet{Shen2012} is (2.7 $\pm$ 0.6) $\times$ 10$^{9} M_{\odot}$. The derived $M_{\rm BH}$/$M_{\rm dyn}$ ratio is 0.020, which is about 4 times higher than the present-day $M_{\rm BH}$/$M_{\rm bulge}$ ratio (0.0051, calculated taking our dynamical mass as the bulge mass by Equation 11 in \citealt{Ho2013}), suggesting that the SMBH grows its mass earlier than the bulge in this luminous quasar at $z$ = 6.13 \citep{Wang2013}.

The dynamical mass would be 21.2 $\times$ 10$^{10}$ $M_{\odot}$ and 8.1 $\times$ 10$^{10}$ $M_{\odot}$ if we adopt the rotational velocities fitted with $i$ = 26 $\degr$ and 48 $\degr$, respectively. And as a sequence, the $M_{\rm BH}$/$M_{\rm dyn}$ ratio would be 0.013 and 0.033, which are 2 and 7 times higher than the local values. Considering these uncertainties, we adopt $M_{\rm dyn}$ = 13.4$_{-5.3}^{+7.8}$ $\times 10^{10}\,M_{\odot}$ and $M_{\rm BH}/M_{\rm dyn}$ = 0.020$_{-0.007}^{+0.013}$  as the final measurements of the dynamical mass and mass ratio.

Note that these results are based on a pure rotation disk model. The dynamical property of the gas component in the nuclear region of such a luminous quasar-starburst system could be more complicated. e.g., \citet{Curtis2016} modeled the feedback from a $z$ $\sim$ 5 quasar and found a rotational star-forming disk and a strong quasar-driven outflow. With the current data of J1319$+$0950, we cannot fully rule out that the velocity gradient is due to a bi-directional outflow, which introduces additional uncertainty of the dynamical mass. Deep observations of the [\ion{C}{2}]-emitting gas at a higher spatial resolution is required to increase data points for detailed dynamical model fit, improve the measurement of the disk inclination angle, and address if there is outflowing gas component in this system.  

\section{Summary}
\label{sec4}

In this paper, we presented ALMA Cycle 1 observations of the dust continuum and [\ion{C}{2}]  line emission in a FIR-luminous quasar J1319$+$0950 at $z=6.13$. Combined with our early ALMA Cycle 0 data, we spatially resolved both the dust continuum and the [\ion{C}{2}] line emission with an angular resolution $\sim$ $0\farcs3$. The [\ion{C}{2}] line emission is more irregular than the dust continuum emission which may suggest difference in their distributions. The flat-peak feature of the [\ion{C}{2}] line spectrum and the clear velocity gradient of the [\ion{C}{2}] velocity map indicate that the gas may be in a rotating disk. We used a tilted ring model to fit the [\ion{C}{2}] velocity field. Our best-fit results yield an inclination angle of 34 $\degr$ and a circular velocity of 427 $\pm$ 55 km s$^{-1}$ at a radius of 3.2 kpc. Finally we calculated a $M_{\rm dyn}$ of 13.4$_{-5.3}^{+7.8}$ $\times 10^{10}\,M_{\odot}$, and a $M_{\rm BH}$/$M_{\rm dyn}$ ratio of 0.020$_{-0.007}^{+0.013}$, which is about 4$_{-2}^{+3}$ times higher than the present-day $M_{\rm BH}$/$M_{\rm bulge}$ ratio. This suggests that in this quasar-starburst system, the SMBH evolves earlier than its bulge in the early evolution phase.

\acknowledgments

This work was supported by National Key Program for Science and Technology Research and Development (grant 2016YFA0400703) and the China Scholarship Council. GCJ is grateful for support from NRAO through the Grote Reber Doctoral Fellowship Program. D.R. acknowledges support from the National Science Foundation under grant number AST-1614213. RW acknowledge supports from the National Science Foundation of China (NSFC) grants No. 11473004, 11533001,  and the National Key Program for Science and Technology Research and Development (grant 2016YFA0400703). We also thank professor D. J. Mortlock for providing the Gemini NIRI spectrum of J1319+0950. This work makes use of the following ALMA data: ADS$/$JAO.ALMA$\#$2011.0.00206.S and ADS$/$JAO.ALMA$\#$2012.1.00240.S. ALMA is a partnership of ESO (representing its member states), NSF (USA) and NINS (Japan), together with NRC (Canada), MOST and ASIAA (Taiwan), and KASI (Republic of Korea), in cooperation with the Republic of Chile. The Joint ALMA Observatory is operated by ESO, AUI$/$NRAO and NAOJ. The National Radio Astronomy Observatory is a facility of the National Science Foundation operated under cooperative agreement by Associated Universities, Inc.

{\it Facilities:} \facility{ALMA}.

\bibliographystyle{apj} 
\bibliography{ms} 

\begin{thebibliography}{}
\expandafter\ifx\csname natexlab\endcsname\relax\def\natexlab#1{#1}\fi

\bibitem[{{Lundgren} {et~al.}(2012)}]{Lundgren2012}
{A. Lundgren}, 2012, ALMA Cycle 1 Technical Handbook, Version 1.01, ALMA

\bibitem[{{Ba{\~n}ados} {et~al.}(2015){Ba{\~n}ados}, {Decarli}, {Walter},
  {Venemans}, {Farina}, \& {Fan}}]{Banados2015}
{Ba{\~n}ados}, E., {Decarli}, R., {Walter}, F., {et~al.} 2015, \apjl, 805, L8

\bibitem[{{Ba{\~n}ados} {et~al.}(2016){Ba{\~n}ados}, {Venemans}, {Decarli},
  {Farina}, {Mazzucchelli}, {Walter}, {Fan}, {Stern}, {Schlafly}, {Chambers},
  {Rix}, {Jiang}, {McGreer}, {Simcoe}, {Wang}, {Yang}, {Morganson}, {De Rosa},
  {Greiner}, {Balokovi{\'c}}, {Burgett}, {Cooper}, {Draper}, {Flewelling},
  {Hodapp}, {Jun}, {Kaiser}, {Kudritzki}, {Magnier}, {Metcalfe}, {Miller},
  {Schindler}, {Tonry}, {Wainscoat}, {Waters}, \& {Yang}}]{Banados2016}
{Ba{\~n}ados}, E., {Venemans}, B.~P., {Decarli}, R., {et~al.} 2016, \apjs, 227,
  11

\bibitem[{{Beelen} {et~al.}(2006){Beelen}, {Cox}, {Benford}, {Dowell},
  {Kov{\'a}cs}, {Bertoldi}, {Omont}, \& {Carilli}}]{Beelen2006}
{Beelen}, A., {Cox}, P., {Benford}, D.~J., {et~al.} 2006, \apj, 642, 694

\bibitem[{{Begeman}(1987)}]{Begeman1987}
{Begeman}, K.~G. 1987, PhD thesis, , Kapteyn Institute, (1987)

\bibitem[{{Bertoldi} {et~al.}(2003{\natexlab{a}}){Bertoldi}, {Carilli}, {Cox},
  {Fan}, {Strauss}, {Beelen}, {Omont}, \& {Zylka}}]{Bertoldi2003a}
{Bertoldi}, F., {Carilli}, C.~L., {Cox}, P., {et~al.} 2003{\natexlab{a}}, \aap,
  406, L55

\bibitem[{{Bertoldi} {et~al.}(2003{\natexlab{b}}){Bertoldi}, {Cox}, {Neri},
  {Carilli}, {Walter}, {Omont}, {Beelen}, {Henkel}, {Fan}, {Strauss}, \&
  {Menten}}]{Bertoldi2003b}
{Bertoldi}, F., {Cox}, P., {Neri}, R., {et~al.} 2003{\natexlab{b}}, \aap, 409,
  L47

\bibitem[{{Bosma}(1978)}]{Bosma1978}
{Bosma}, A. 1978, PhD thesis, PhD Thesis, Groningen Univ., (1978)

\bibitem[{{Bothwell} {et~al.}(2010){Bothwell}, {Chapman}, {Tacconi}, {Smail},
  {Ivison}, {Casey}, {Bertoldi}, {Beswick}, {Biggs}, {Blain}, {Cox}, {Genzel},
  {Greve}, {Kennicutt}, {Muxlow}, {Neri}, \& {Omont}}]{Bothwell2010}
{Bothwell}, M.~S., {Chapman}, S.~C., {Tacconi}, L., {et~al.} 2010, \mnras, 405,
  219

\bibitem[{{Bouch{\'e}} {et~al.}(2007){Bouch{\'e}}, {Cresci}, {Davies},
  {Eisenhauer}, {F{\"o}rster Schreiber}, {Genzel}, {Gillessen}, {Lehnert},
  {Lutz}, {Nesvadba}, {Shapiro}, {Sternberg}, {Tacconi}, {Verma}, {Cimatti},
  {Daddi}, {Renzini}, {Erb}, {Shapley}, \& {Steidel}}]{Bouche2007}
{Bouch{\'e}}, N., {Cresci}, G., {Davies}, R., {et~al.} 2007, \apj, 671, 303

\bibitem[{{Chabrier}(2003)}]{Chabrier2003}
{Chabrier}, G. 2003, \apjl, 586, L133

\bibitem[{{Curtis} \& {Sijacki}(2016)}]{Curtis2016}
{Curtis}, M., \& {Sijacki}, D. 2016, \mnras, 457, L34

\bibitem[{{Daddi} {et~al.}(2010){Daddi}, {Bournaud}, {Walter}, {Dannerbauer},
  {Carilli}, {Dickinson}, {Elbaz}, {Morrison}, {Riechers}, {Onodera}, {Salmi},
  {Krips}, \& {Stern}}]{Daddi2010}
{Daddi}, E., {Bournaud}, F., {Walter}, F., {et~al.} 2010, \apj, 713, 686

\bibitem[{{de Blok} \& {McGaugh}(1997)}]{deBlok1997}
{de Blok}, W.~J.~G., \& {McGaugh}, S.~S. 1997, \mnras, 290, 533

\bibitem[{{D{\'{\i}}az-Santos} {et~al.}(2016){D{\'{\i}}az-Santos}, {Assef},
  {Blain}, {Tsai}, {Aravena}, {Eisenhardt}, {Wu}, {Stern}, \&
  {Bridge}}]{DiazSantos2016}
{D{\'{\i}}az-Santos}, T., {Assef}, R.~J., {Blain}, A.~W., {et~al.} 2016, \apjl,
  816, L6

\bibitem[{{Fall} \& {Efstathiou}(1980)}]{Fall1980}
{Fall}, S.~M., \& {Efstathiou}, G. 1980, \mnras, 193, 189

\bibitem[{{Fan} {et~al.}(2006){Fan}, {Carilli}, \& {Keating}}]{Fan2006lala}
{Fan}, X., {Carilli}, C.~L., \& {Keating}, B. 2006, \araa, 44, 415

\bibitem[{{Hodge} {et~al.}(2015){Hodge}, {Riechers}, {Decarli}, {Walter},
  {Carilli}, {Daddi}, \& {Dannerbauer}}]{Hodge2015}
{Hodge}, J.~A., {Riechers}, D., {Decarli}, R., {et~al.} 2015, \apjl, 798, L18

\bibitem[{{Jiang} {et~al.}(2015){Jiang}, {McGreer}, {Fan}, {Bian}, {Cai},
  {Cl{\'e}ment}, {Wang}, \& {Fan}}]{Jiang2015}
{Jiang}, L., {McGreer}, I.~D., {Fan}, X., {et~al.} 2015, \aj, 149, 188

\bibitem[{{Jiang} {et~al.}(2016){Jiang}, {McGreer}, {Fan}, {Strauss},
  {Ba{\~n}ados}, {Becker}, {Bian}, {Farnsworth}, {Shen}, {Wang}, {Wang},
  {Wang}, {White}, {Wu}, {Wu}, {Yang}, \& {Yang}}]{Jiang2016}
---. 2016, \apj, 833, 222

\bibitem[{{Kamphuis} {et~al.}(2015){Kamphuis}, {J{\'o}zsa}, {Oh}, {Spekkens},
  {Urbancic}, {Serra}, {Koribalski}, \& {Dettmar}}]{Kamphuis2015}
{Kamphuis}, P., {J{\'o}zsa}, G.~I.~G., {Oh}, S.-.~H., {et~al.} 2015, \mnras,
  452, 3139

\bibitem[{{Kennicutt}(1998{\natexlab{a}})}]{Kennicutt1998}
{Kennicutt}, Jr., R.~C. 1998{\natexlab{a}}, \araa, 36, 189

\bibitem[{{Kennicutt}(1998{\natexlab{b}})}]{Kennicutt1998sl}
---. 1998{\natexlab{b}}, \apj, 498, 541

\bibitem[{{Kimball} {et~al.}(2015){Kimball}, {Lacy}, {Lonsdale}, \&
  {Macquart}}]{Kimball2015}
{Kimball}, A.~E., {Lacy}, M., {Lonsdale}, C.~J., \& {Macquart}, J.-P. 2015,
  \mnras, 452, 88

\bibitem[{{Kormendy} \& {Ho}(2013)}]{Ho2013}
{Kormendy}, J., \& {Ho}, L.~C. 2013, \araa, 51, 511

\bibitem[{{Lucero} {et~al.}(2015){Lucero}, {Carignan}, {Elson},
  {Randriamampandry}, {Jarrett}, {Oosterloo}, \& {Heald}}]{Lucero2015}
{Lucero}, D.~M., {Carignan}, C., {Elson}, E.~C., {et~al.} 2015, \mnras, 450,
  3935

\bibitem[{{Maiolino} {et~al.}(2012){Maiolino}, {Gallerani}, {Neri}, {Cicone},
  {Ferrara}, {Genzel}, {Lutz}, {Sturm}, {Tacconi}, {Walter}, {Feruglio},
  {Fiore}, \& {Piconcelli}}]{Maiolino2012}
{Maiolino}, R., {Gallerani}, S., {Neri}, R., {et~al.} 2012, \mnras, 425, L66

\bibitem[{{Matsuoka} {et~al.}(2016){Matsuoka}, {Onoue}, {Kashikawa}, {Iwasawa},
  {Strauss}, {Nagao}, {Imanishi}, {Niida}, {Toba}, {Akiyama}, {Asami}, {Bosch},
  {Foucaud}, {Furusawa}, {Goto}, {Gunn}, {Harikane}, {Ikeda}, {Kawaguchi},
  {Kikuta}, {Komiyama}, {Lupton}, {Minezaki}, {Miyazaki}, {Morokuma},
  {Murayama}, {Nishizawa}, {Ono}, {Ouchi}, {Price}, {Sameshima}, {Silverman},
  {Sugiyama}, {Tait}, {Takada}, {Takata}, {Tanaka}, {Tang}, \&
  {Utsumi}}]{Matsuoka2016}
{Matsuoka}, Y., {Onoue}, M., {Kashikawa}, N., {et~al.} 2016, \apj, 828, 26

\bibitem[McGaugh et al.(2001)]{McGaugh2001} McGaugh, S.~S., Rubin, V.~C., \& de Blok, W.~J.~G.\ 2001, \aj, 122, 2381

\bibitem[{{Mortlock} {et~al.}(2011){Mortlock}, {Warren}, {Patel}, {Venemans},
  {McMahon}, {Hewett}, {Simpson}, {Theuns}, {Rottgering}, {Kuiper}, {Bolton},
  \& {Harhnelt}}]{Mortlock2011}
{Mortlock}, D., {Warren}, S., {Patel}, M., {et~al.} 2011, in Galaxy Formation,
  88

\bibitem[{{Mortlock} {et~al.}(2009){Mortlock}, {Patel}, {Warren}, {Venemans},
  {McMahon}, {Hewett}, {Simpson}, {Sharp}, {Burningham}, {Dye}, {Ellis},
  {Gonzales-Solares}, \& {Hu{\'e}lamo}}]{Mortlock2009}
{Mortlock}, D.~J., {Patel}, M., {Warren}, S.~J., {et~al.} 2009, \aap, 505, 97

\bibitem[{{Petric} {et~al.}(2003){Petric}, {Carilli}, {Bertoldi}, {Fan}, {Cox},
  {Strauss}, {Omont}, \& {Schneider}}]{Petric2003}
{Petric}, A.~O., {Carilli}, C.~L., {Bertoldi}, F., {et~al.} 2003, \aj, 126, 15

\bibitem[{{Priddey} {et~al.}(2003){Priddey}, {Isaak}, {McMahon}, {Robson}, \&
  {Pearson}}]{Priddey2003}
{Priddey}, R.~S., {Isaak}, K.~G., {McMahon}, R.~G., {Robson}, E.~I., \&
  {Pearson}, C.~P. 2003, \mnras, 344, L74

\bibitem[{{Reed} {et~al.}(2017){Reed}, {McMahon}, {Martini}, {Banerji},
  {Auger}, {Hewett}, {Koposov}, {Gibbons}, {Gonzalez-Solares}, {Ostrovski},
  {Tie}, {Abdalla}, {Allam}, {Benoit-Levy}, {Bertin}, {Brooks}, {Buckley-Geer},
  {Burke}, {Carnero Rosell}, {Carrasco Kind}, {Carretero}, {da Costa}, {DePoy},
  {Desai}, {Diehl}, {Doel}, {Evrard}, {Finley}, {Flaugher}, {Fosalba},
  {Frieman}, {Garc{\i}a-Bellido}, {Gaztanaga}, {Goldstein}, {Gruen}, {Gruendl},
  {Gutierrez}, {James}, {Kuehn}, {Kuropatkin}, {Lahav}, {Lima}, {Maia},
  {Marshall}, {Melchior}, {Miller}, {Miquel}, {Nord}, {Ogando}, {Plazas},
  {Romer}, {Sanchez}, {Scarpine}, {Schubnell}, {Sevilla-Noarbe}, {Smith},
  {Sobreira}, {Suchyta}, {Swanson}, {Tarle}, {Tucker}, {Walker}, \&
  {Wester}}]{Reed2017}
{Reed}, S.~L., {McMahon}, R.~G., {Martini}, P., {et~al.} 2017, ArXiv e-prints,
  arXiv:1701.04852

\bibitem[{{Riechers} {et~al.}(2009){Riechers}, {Walter}, {Bertoldi}, {Carilli},
  {Aravena}, {Neri}, {Cox}, {Wei{\ss}}, \& {Menten}}]{Riechers2009}
{Riechers}, D.~A., {Walter}, F., {Bertoldi}, F., {et~al.} 2009, \apj, 703, 1338

\bibitem[{{Rogstad} {et~al.}(1974){Rogstad}, {Lockhart}, \&
  {Wright}}]{Rogstad1974}
{Rogstad}, D.~H., {Lockhart}, I.~A., \& {Wright}, M.~C.~H. 1974, \apj, 193, 309

\bibitem[{{Shen} \& {Liu}(2012)}]{Shen2012}
{Shen}, Y., \& {Liu}, X. 2012, \apj, 753, 125

\bibitem[{{Spergel} {et~al.}(2007){Spergel}, {Bean}, {Dor{\'e}}, {Nolta},
  {Bennett}, {Dunkley}, {Hinshaw}, {Jarosik}, {Komatsu}, {Page}, {Peiris},
  {Verde}, {Halpern}, {Hill}, {Kogut}, {Limon}, {Meyer}, {Odegard}, {Tucker},
  {Weiland}, {Wollack}, \& {Wright}}]{Spergel2007}
{Spergel}, D.~N., {Bean}, R., {Dor{\'e}}, O., {et~al.} 2007, \apjs, 170, 377

\bibitem[Swaters et al.(2000)]{Swaters2000} Swaters, R.~A., Madore, B.~F., \& Trewhella, M.\ 2000, \apjl, 531, L107

\bibitem[{{Swaters} {et~al.}(2009){Swaters}, {Sancisi}, {van Albada}, \& {van
  der Hulst}}]{Swaters2009}
{Swaters}, R.~A., {Sancisi}, R., {van Albada}, T.~S., \& {van der Hulst}, J.~M.
  2009, \aap, 493, 871

\bibitem[{{Tacconi} {et~al.}(2010){Tacconi}, {Genzel}, {Neri}, {Cox}, {Cooper},
  {Shapiro}, {Bolatto}, {Bouch{\'e}}, {Bournaud}, {Burkert}, {Combes},
  {Comerford}, {Davis}, {Schreiber}, {Garcia-Burillo}, {Gracia-Carpio}, {Lutz},
  {Naab}, {Omont}, {Shapley}, {Sternberg}, \& {Weiner}}]{Tacconi2010}
{Tacconi}, L.~J., {Genzel}, R., {Neri}, R., {et~al.} 2010, \nat, 463, 781

\bibitem[{{van der Hulst} {et~al.}(1992){van der Hulst}, {Terlouw}, {Begeman},
  {Zwitser}, \& {Roelfsema}}]{vanderHulst1992}
{van der Hulst}, J.~M., {Terlouw}, J.~P., {Begeman}, K.~G., {Zwitser}, W., \&
  {Roelfsema}, P.~R. 1992, in Astronomical Society of the Pacific Conference
  Series, Vol.~25, Astronomical Data Analysis Software and Systems I, ed. D.~M.
  {Worrall}, C.~{Biemesderfer}, \& J.~{Barnes}, 131

\bibitem[{{Venemans} {et~al.}(2007){Venemans}, {McMahon}, {Warren},
  {Gonzalez-Solares}, {Hewett}, {Mortlock}, {Dye}, \& {Sharp}}]{Venemans2007}
{Venemans}, B.~P., {McMahon}, R.~G., {Warren}, S.~J., {et~al.} 2007, \mnras,
  376, L76

\bibitem[{{Venemans} {et~al.}(2016){Venemans}, {Walter}, {Zschaechner},
  {Decarli}, {De Rosa}, {Findlay}, {McMahon}, \& {Sutherland}}]{Venemans2016}
{Venemans}, B.~P., {Walter}, F., {Zschaechner}, L., {et~al.} 2016, \apj, 816,
  37

\bibitem[{{Venemans} {et~al.}(2012){Venemans}, {McMahon}, {Walter}, {Decarli},
  {Cox}, {Neri}, {Hewett}, {Mortlock}, {Simpson}, \& {Warren}}]{Venemans2012}
{Venemans}, B.~P., {McMahon}, R.~G., {Walter}, F., {et~al.} 2012, \apjl, 751,
  L25

\bibitem[{{Venemans} {et~al.}(2015{\natexlab{a}}){Venemans}, {Verdoes Kleijn},
  {Mwebaze}, {Valentijn}, {Ba{\~n}ados}, {Decarli}, {de Jong}, {Findlay},
  {Kuijken}, {Barbera}, {McFarland}, {McMahon}, {Napolitano}, {Sikkema}, \&
  {Sutherland}}]{Venemans2015a}
{Venemans}, B.~P., {Verdoes Kleijn}, G.~A., {Mwebaze}, J., {et~al.}
  2015{\natexlab{a}}, \mnras, 453, 2259

\bibitem[{{Venemans} {et~al.}(2015{\natexlab{b}}){Venemans}, {Ba{\~n}ados},
  {Decarli}, {Farina}, {Walter}, {Chambers}, {Fan}, {Rix}, {Schlafly},
  {McMahon}, {Simcoe}, {Stern}, {Burgett}, {Draper}, {Flewelling}, {Hodapp},
  {Kaiser}, {Magnier}, {Metcalfe}, {Morgan}, {Price}, {Tonry}, {Waters},
  {AlSayyad}, {Banerji}, {Chen}, {Gonz{\'a}lez-Solares}, {Greiner},
  {Mazzucchelli}, {McGreer}, {Miller}, {Reed}, \& {Sullivan}}]{Venemans2015b}
{Venemans}, B.~P., {Ba{\~n}ados}, E., {Decarli}, R., {et~al.}
  2015{\natexlab{b}}, \apjl, 801, L11

\bibitem[{{Venemans} {et~al.}(2017){Venemans}, {Walter}, {Decarli},
  {Ba{\~n}ados}, {Hodge}, {Hewett}, {McMahon}, {Mortlock}, \&
  {Simpson}}]{Venemans2017}
{Venemans}, B.~P., {Walter}, F., {Decarli}, R., {et~al.} 2017, \apj, 837, 146

\bibitem[{{Walter} {et~al.}(2009){Walter}, {Riechers}, {Cox}, {Neri},
  {Carilli}, {Bertoldi}, {Weiss}, \& {Maiolino}}]{Walter2009}
{Walter}, F., {Riechers}, D., {Cox}, P., {et~al.} 2009, \nat, 457, 699

\bibitem[{{Wang} {et~al.}(2008){Wang}, {Carilli}, {Wagg}, {Bertoldi}, {Walter},
  {Menten}, {Omont}, {Cox}, {Strauss}, {Fan}, {Jiang}, \&
  {Schneider}}]{Wang2008thermal}
{Wang}, R., {Carilli}, C.~L., {Wagg}, J., {et~al.} 2008, \apj, 687, 848

\bibitem[{{Wang} {et~al.}(2011){Wang}, {Wagg}, {Carilli}, {Neri}, {Walter},
  {Omont}, {Riechers}, {Bertoldi}, {Menten}, {Cox}, {Strauss}, {Fan}, \&
  {Jiang}}]{Wang2011fir}
{Wang}, R., {Wagg}, J., {Carilli}, C.~L., {et~al.} 2011, \aj, 142, 101

\bibitem[{{Wang} {et~al.}(2013){Wang}, {Wagg}, {Carilli}, {Walter}, {Lentati},
  {Fan}, {Riechers}, {Bertoldi}, {Narayanan}, {Strauss}, {Cox}, {Omont},
  {Menten}, {Knudsen}, {Neri}, \& {Jiang}}]{Wang2013}
---. 2013, \apj, 773, 44

\bibitem[{{Wang} {et~al.}(2016){Wang}, {Wu}, {Neri}, {Fan}, {Walter},
  {Carilli}, {Momjian}, {Bertoldi}, {Strauss}, {Li}, {Wang}, {Riechers},
  {Jiang}, {Omont}, {Wagg}, \& {Cox}}]{WangRan2016}
{Wang}, R., {Wu}, X.-B., {Neri}, R., {et~al.} 2016, \apj, 830, 53

\bibitem[{{Willott} {et~al.}(2015){Willott}, {Bergeron}, \&
  {Omont}}]{Willott2015}
{Willott}, C.~J., {Bergeron}, J., \& {Omont}, A. 2015, \apj, 801, 123

\bibitem[{{Willott} {et~al.}(2013){Willott}, {Omont}, \&
  {Bergeron}}]{Willott2013}
{Willott}, C.~J., {Omont}, A., \& {Bergeron}, J. 2013, \apj, 770, 13

\end{thebibliography}

\end{document}